\documentclass[iop]{emulateapj}
\RequirePackage{rotating}
\usepackage[]{hyperref,epstopdf,amsmath,xcolor}
\usepackage[]{empheq}
\usepackage{ccaption}

\slugcomment{Draft \today}

\shorttitle{Do non-dipolar magnetic fields contribute to spin-down torques?}
\shortauthors{V. See et al.}

\begin{document}

\title{Do non-dipolar magnetic fields contribute to spin-down torques?}

\author{Victor See$^{1}$, Sean P. Matt$^{1}$, Adam J. Finley$^{1}$, Colin P. Folsom$^{2,3}$,\\
Sudeshna Boro Saikia$^{4}$, Jean-Francois Donati$^{2,3}$, Rim Fares$^{5}$, \'{E}lodie M. H\'{e}brard, \\
Moira M. Jardine$^{6}$, Sandra V. Jeffers$^{7}$, Stephen C. Marsden$^{8}$, Matthew W. Mengel$^{8}$, \\
Julien Morin$^{9}$, Pascal Petit$^{2,3}$, Aline A. Vidotto$^{10}$, Ian A. Waite$^{8}$ and the BCool collaboration}
\affil{$^{1}$University of Exeter, Deparment of Physics \& Astronomy, Stocker Road, Devon, Exeter, EX4 4QL, UK\\
$^{2}$Universit\'{e} de Toulouse, UPS-OMP, IRAP, Toulouse, France\\
$^{3}$CNRS, Institut de Recherche en Astrophysique et Planetologie, 14, avenue Edouard Belin, F-31400 Toulouse, France\\
$^{4}$University of Vienna, Department of Astrophysics, Türkenschanzstrasse 17, 1180 Vienna, Austria\\
$^{5}$Physics Department, United Arab Emirates University, P.O. Box 15551, Al-Ain, United Arab Emirates\\
$^{6}$SUPA, School of Physics and Astronomy, University of St Andrews, North Haugh, St Andrews KY16 9SS, UK\\
$^{7}$Universit\"{a}t G\"{o}ttingen, Institut f\"{u}r Astrophysik, Friedrich-Hund-Platz 1, D-37077 G\"{o}ttingen, Germany\\
$^{8}$University of Southern Queensland, Centre for Astrophysics, Toowoomba, QLD 4350, Australia\\
$^{9}$Laboratoire Univers et Particules de Montpellier, Universit\'{e} ́ de Montpellier, CNRS, F-34095, France\\
$^{10}$School of Physics, Trinity College Dublin, University of Dublin, Dublin-2, Ireland
}
\email{*w.see@exeter.ac.uk}

\begin{abstract}
Main sequence low-mass stars are known to spin-down as a consequence of their magnetised stellar winds. However, estimating the precise rate of this spin-down is an open problem. The mass-loss rate, angular momentum-loss rate and the magnetic field properties of low-mass stars are fundamentally linked making this a challenging task. Of particular interest is the stellar magnetic field geometry. In this work, we consider whether non-dipolar field modes contribute significantly to the spin-down of low-mass stars. We do this using a sample of stars that have all been previously mapped with Zeeman-Doppler imaging. For a given star, as long as its mass-loss rate is below some critical mass-loss rate, only the dipolar fields contribute to its spin-down torque. However, if it has a larger mass-loss rate, higher order modes need to be considered. For each star, we calculate this critical mass-loss rate, which is a simple function of the field geometry. Additionally, we use two methods of estimating mass-loss rates for our sample of stars. In the majority of cases, we find that the estimated mass-loss rates do not exceed the critical mass-loss rate and hence, the dipolar magnetic field alone is sufficient to determine the spin-down torque. However, we find some evidence that, at large Rossby numbers, non-dipolar modes may start to contribute.
\end{abstract}


\section{Introduction}
\label{sec:Intro}
It is well known that thermally driven stellar winds cause low-mass stars to spin down over their main sequence lifetimes \citep[e.g.][]{Weber1967}. Many authors have attempted to model the rotation period evolution of these stars \citep{Gallet2013,Brown2014,Gallet2015,Matt2015,Johnstone2015,Amard2016,Blackman2016,vanSaders2016,Gondoin2017,Ardestani2017,See2018,Garraffo2018} but accurately determining angular momentum-loss rates is a difficult task. An open question in this context is the impact of magnetic field geometry. Parameter studies using MHD simulations have shown that, when considering only single spherical harmonic modes, the angular momentum-loss rate is highest when the field is dipolar and drops dramatically for higher order field configurations, e.g. quadrupolar, octupolar, etc. \citep{Garraffo2015,Reville2015}. However, the magnetic fields of real stars are known to be a mixture of many spherical harmonic modes \citep[e.g.][]{DeRosa2012}. It turns out that, when constructing so-called ``braking laws'' that  specify the rate at which stars lose angular momentum, the relevant magnetic parameter to consider is the open flux \citep{Vidotto2014Torque,Reville2015,Finley2017,Pantolmos2017,Finley2018}. This is the flux associated with field lines that extend into interplanetary space, i.e. the field lines along which stellar winds carry away mass and angular momentum.

Previous studies have suggested that the open flux, and hence angular momentum-loss rates, are dominated by the dipolar component of the stellar magnetic field \citep{Jardine2017,See2017,See2018}. However, the open flux is not a directly observable quantity and attempts to estimate it are difficult and model dependent. Indeed, there have been some suggestions that higher order magnetic field modes may play a significant role in the spin-down of low-mass stars \citep{Brown2014,Garraffo2018}. Recently, \citet{Finley2017,Finley2018} formulated a braking law in terms of the photospheric field strengths of the dipole, quadrupole and octupole components of the field. Formulating braking laws in this manner is advantageous because the field geometry can be accounted for without resorting to model dependent open flux estimates. This braking law can therefore be used to more precisely test the claim that dipole magnetic fields dominate over higher order geometries when estimating angular momentum-loss rates.

To use the braking law of \citet[][henceforth F18]{Finley2018}, we have to determine the field strengths associated with the dipole, quadrupole and octupole components of stellar magnetic fields. This is something that the Zeeman-Doppler imaging (ZDI) technique can uniquely do. ZDI is a tomographic technique that can reconstruct the large-scale photospheric magnetic field geometries of low-mass stars \citep{Semel1989,Brown1991,Donati1997,Donati2006}. The magnetic field maps produced from ZDI are expressed in terms of a spherical harmonic decomposition (e.g. see appendix B of \citet{Folsom2018}) allowing us to take advantage of the F18 braking law.

As well as the magnetic properties of the star, knowledge of the mass-loss rate is also important to accurately determine angular momentum-loss rates. Mass-loss rates for low-mass stars are notoriously difficult to determine due to the diffuse nature of their winds. Currently, only indirect methods of measuring stellar mass-loss rates are possible \citep{Wood2014,Vidotto2017,Jardine2018}. Consequently, when using braking laws, the mass-loss rate is likely to be the least well constrained input parameter.  

In this paper, we investigate whether the dipole magnetic field dominates when calculating angular-momentum losses using the F18 braking law. In section \ref{sec:torque}, we present this braking law and demonstrate that only the dipole component of the magnetic field contributes to a stars's spin-down torque if the star's mass-loss rate is below some critical mass-loss rate. In section \ref{sec:MagProps}, we discuss the magnetic properties of the ZDI sample used in this work. Two methods for estimating mass-loss rates are introduced in section \ref{sec:Mdots}. In section \ref{sec:CriticalMdots}, we calculate the critical mass-loss rates for our sample and analyse which stars are the most likely to have mass-loss rates that exceed the critical mass-loss rate. A discussion of the uncertainties of our work are presented in section \ref{sec:Uncertainties}. Finally, a discussion and the conclusions of our results are presented in section \ref{sec:Conclusions}.

\section{Finley \& Matt (2018) braking law}
\label{sec:torque}
The braking law of F18 takes the form of a twice broken power law and is given by 

\begin{equation}
	T = \dot{M}\Omega_{\star}\langle R_{\rm A}\rangle^2
	\label{eq:Torque}
\end{equation}
where $T$ is the angular momentum loss-rate or spin-down torque, $\dot{M}$ is the mass-loss rate, $\Omega_{\star}=2\pi/P_{\rm rot}$ is the angular frequency, $P_{\rm rot}$ is the rotation period and $\langle R_{\rm A}\rangle$ is the torque averaged Alfv\'{e}n radius. Physically, $\langle R_{\rm A}\rangle$ corresponds to the lever arm of the spin-down torque and is given by

\begin{subequations}
\label{eq:BrakingLaw}
    \begin{empheq}[left={\langle R_{\rm A}\rangle/r_{\star}={\rm max}\empheqlbrace\,}]{align}
      & K_{\rm d}\left[\mathcal{R}_{\rm d}^2\Upsilon\right]^{m_{\rm d}}
      \label{eq:BrakingLawDip} \\
      & K_{\rm q}\left[\left(\mathcal{R}_{\rm d}+\mathcal{R}_{\rm q}\right)^2\Upsilon\right]^{m_{\rm q}}
      \label{eq:BrakingLawQuad} \\
      & K_{\rm o}\left[\left(\mathcal{R}_{\rm d}+\mathcal{R}_{\rm q}+\mathcal{R}_{\rm o}\right)^2\Upsilon\right]^{m_{\rm o}}.
      \label{eq:BrakingLawOct}
    \end{empheq}
\end{subequations}
Here, $r_{\star}$ is the stellar radius, $\Upsilon = \frac{B_{\star}^2 r_{\star}^2}{\dot{M}v_{\rm esc}}$ is the wind magnetisation and $v_{\rm esc}$ is the escape velocity of the star. $\mathcal{R}_{\rm d} = B_{\rm d}/B_{\star}$, $\mathcal{R}_{\rm q} = B_{\rm q}/B_{\star}$ and $\mathcal{R}_{\rm o} = B_{\rm o}/B_{\star}$ are the magnetic field ratios where $B_{\star} = B_{\rm d} + B_{\rm q} + B_{\rm o}$. By definition, $\mathcal{R}_{\rm d} + \mathcal{R}_{\rm q} + \mathcal{R}_{\rm o} = 1$. The subscripts $\rm d$, $\rm q$ and $\rm o$ indicate dipole, quadrupole and octopole respectively. The precise meaning of $B_{\rm d}$, $B_{\rm q}$ and $B_{\rm o}$ will be further discussed in section \ref{sec:MagProps}. Finally, $K_{\rm d} = 1.53$, $K_{\rm q} = 1.7$, $K_{\rm o} = 1.8$, $m_{\rm d} = 0.229$, $m_{\rm q} = 0.134$ and $m_{\rm o} = 0.087$ are fit parameters obtained from the MHD simulations of F18. 

Mathematically, equation (\ref{eq:BrakingLaw}) can predict arbitrarily small Alfv\'{e}n radii if $\Upsilon$ is very small. However, we impose a lower limit on the Alfv\'{e}n radius of $\sqrt{2/3}\ r_{\star}$. This is the lever arm associated with a completely unmagnetised and inviscid wind and smaller values are unphysical. The factor of $\sqrt{2/3}$ is a consequence of our definition of the angular momentum-loss rate in equation (\ref{eq:Torque}). It is a geometric factor that arises when integrating the angular momentum-loss rate over all latitudes (e.g. compare with \citet{Weber1967}). We also note that this braking law only accounts for contributions from the dipolar, quadrupolar and octupolar field modes. Although real stellar magnetic fields contain higher order spherical harmonic modes, this braking law likely provides a reasonable estimate of the torque. As discussed later in this section, high order field modes only become relevant at high mass-loss rates. However, if the mass-loss rate of a star is high enough for field modes above the octupolar mode to contribute, the Alfv\'en radius is likely to be so small that it is close to, or below, the unmagnetised $r_{\rm A}=\sqrt{2/3}\ r_{\star}$ limit. 

\begin{figure}
	\begin{center}
	\includegraphics[trim=0cm 1cm 1cm 0cm,width=\columnwidth]{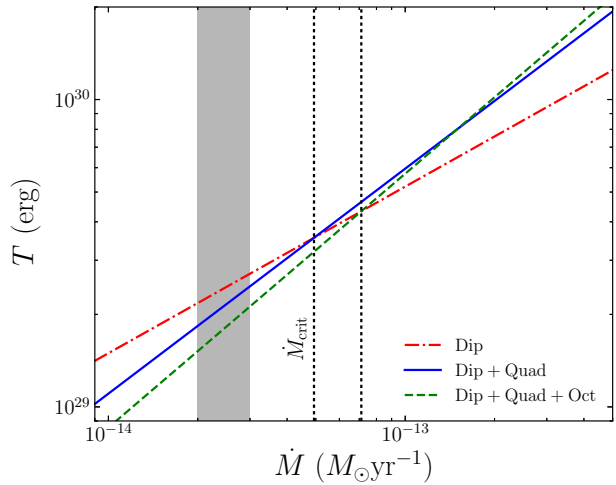}
	\end{center}
	\caption{Angular momentum-loss rate as a function of mass-loss rate using equations (\ref{eq:Torque}) and (\ref{eq:BrakingLaw}) for solar parameters (see table \ref{tab:Sun}). The three power laws show the three components of equation (\ref{eq:BrakingLaw}); red dot-dashed line for the dipole component (equation (\ref{eq:BrakingLawDip})), blue solid line for the dipole + quadrupole component (equation (\ref{eq:BrakingLawQuad})) and green dashed line for the dipole + quadrupole + octupole component (equation (\ref{eq:BrakingLawOct})). The true angular momentum-loss rate is the upper envelope of these 3 power laws. Black dotted lines indicate the mass-loss rate at which equations (\ref{eq:BrakingLawDip}) \& (\ref{eq:BrakingLawQuad}) intersect (labelled as $\dot{M}_{\rm crit}$) and where equations (\ref{eq:BrakingLawDip}) \& (\ref{eq:BrakingLawOct}) intersect. The shaded region roughly indicates the observed solar mass-loss rate.}
	\label{fig:SolarBrakingLaw}
\end{figure}

It is instructive to analyse the behaviour of this braking law. As an illustrative example, in fig. \ref{fig:SolarBrakingLaw} we plot angular momentum-loss rate as a function of mass-loss rate using equations (\ref{eq:Torque}) and (\ref{eq:BrakingLaw}) for solar parameters (these are listed in table \ref{tab:Sun}; the values used for $B_{\rm d, \odot}$, $B_{\rm q, \odot}$ \& $B_{\rm o, \odot}$ are discussed in section \ref{sec:MagProps}). While the mass-loss rate of the Sun is relatively well constrained from observations, the same is not true for other stars. Understanding how this braking law behaves as a function of mass-loss rate is therefore a useful exercise. The three components of equation (\ref{eq:BrakingLaw}) are plotted as three separate power laws. The predicted angular momentum-loss rate from the F18 braking law is the upper envelope of these three functions.

\begin{table}
\begin{minipage}{88mm}
	\begin{center}	
	\caption{Adopted solar parameters. Note that the upper and lower bounds on the field strengths are the range of values observed in cycle 24 rather than formal errors.}
	\label{tab:Sun}
	\begin{tabular}{lcc}
	\hline
	 & Symbol & Adopted value\\
	\hline
	Mass & $M_{\odot}$ & $1.99\times  10^{33}$g\\
	Radius & $r_{\odot}$ & $6.96\times 10^{10}$cm\\
	Angular frequency & $\Omega_{\odot}$ & $2.6\times 10^{-6}$Hz\\
	Dipole field strength & $B_{\rm d, \odot}$ & $0.9^{+0.7}_{-0.6}$G\\
	Quadrupole field strength & $B_{\rm q, \odot}$ & $0.8^{+1.8}_{-0.6}$G\\
	Octupole field strength & $B_{\rm o, \odot}$ & $1.2^{+1.2}_{-0.8}$G\\
	\hline
\end{tabular}
\end{center}
\end{minipage}
\end{table}

Fig. \ref{fig:SolarBrakingLaw} shows that, equation (\ref{eq:BrakingLawDip}; red dot-dashed line) dominates at low mass-loss rates, equation (\ref{eq:BrakingLawQuad}; blue solid line) dominates at intermediate mass-loss rates and equation (\ref{eq:BrakingLawOct}; green dashed line) dominates at high mass-loss rates. Since equation (\ref{eq:BrakingLawDip}) depends only on $B_{\rm d}$, equation (\ref{eq:BrakingLawQuad}) depends on $B_{\rm d}$ \& $B_{\rm q}$ and equation (\ref{eq:BrakingLawOct}) depends on $B_{\rm d}$, $B_{\rm q}$ \& $B_{\rm o}$, the general trend is that more high order, non-dipolar, field modes are required to properly calculate the spin-down torque at higher mass-loss rates. Physically, this is because high order field modes decay more rapidly as a function of distance from the star \citep[e.g. see][]{Reville2015}. If the mass-loss rate is low, the Alfv\'{e}n radius (the lever arm of the spin-down torque) will be large and extend out to a distance where non-dipolar field components have decayed away. However, if the mass-loss rate is large, the Alfv\'{e}n radius may be small enough that high order field components have not completely decayed away and must be accounted for.

We can define a critical mass-loss rate, $\dot{M}_{\rm crit}$, below which only the dipole field strength is required to properly determine the angular momentum-loss rate of the star. This critical mass-loss rate is given by equating equations (\ref{eq:BrakingLawDip}) \& (\ref{eq:BrakingLawQuad}) and solving for the mass-loss rate\footnote{In principle, to properly calculate $\dot{M}_{\rm crit}$, one should calculate the mass-loss rate at which equation (\ref{eq:BrakingLawDip}) equals equation (\ref{eq:BrakingLawQuad}) and also the mass-loss rate at which equation (\ref{eq:BrakingLawDip}) equals equation (\ref{eq:BrakingLawOct}). $\dot{M}_{\rm crit}$ is then given by the smaller of these two mass-loss rates. However, for the sample of stars presented in section \ref{sec:MagProps}, it turns out that the mass-loss rate at which (\ref{eq:BrakingLawDip}) equals equation (\ref{eq:BrakingLawQuad}) is always smaller than the mass-loss rate at which (\ref{eq:BrakingLawDip}) equals equation (\ref{eq:BrakingLawOct}).}. It is given by

\begin{equation}
	\dot{M}_{\rm crit} = 0.33\frac{B_{\star}^2 r_{\star}^2}{v_{\rm esc}} \frac{\mathcal{R}_{\rm d}^{4.82}}{(\mathcal{R}_{\rm d}+\mathcal{R}_{\rm q})^{2.82}},
	\label{eq:MDotCrit}
\end{equation}
where we have substituted in the values for $K_{\rm d}$, $K_{\rm q}$, $m_{\rm d}$ and $m_{\rm q}$. For solar parameters, $\dot{M}_{\rm crit} \sim 5\times10^{-14} M_{\odot}{\rm yr}^{-1}$. The observed mass-loss rate for the Sun, $\dot{M} \sim 2\times10^{-14} M_{\odot}{\rm yr}^{-1}$, is lower than this critical mass-loss rate and, consequently, we only need the dipole component of the solar magnetic field to calculate the solar angular-momentum loss rate using the F18 braking law. However, there are caveats to this statement, related to the variability of the Sun's magnetic activity, that we shall return to in section \ref{subsec:Variability}.

\begin{figure*}
	\begin{center}
	\includegraphics[trim=1cm 1.5cm 1cm 0cm,width=\textwidth]{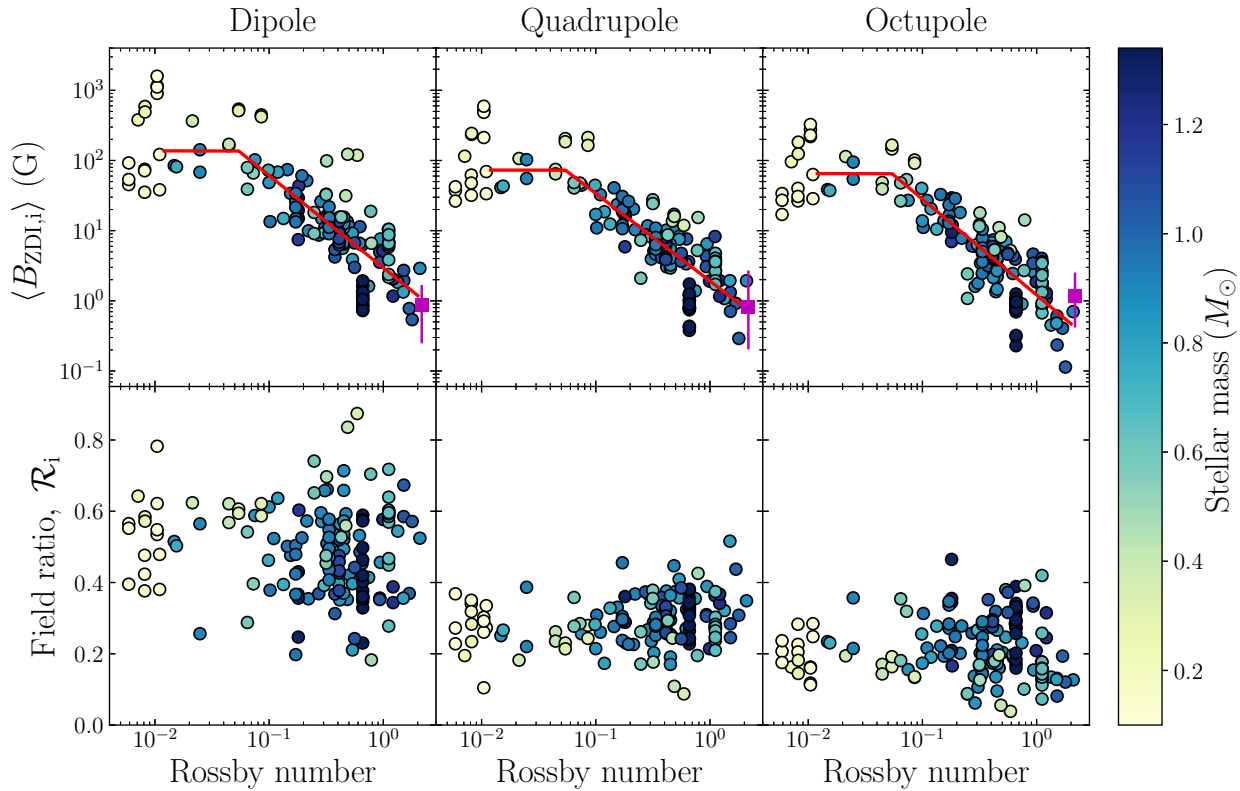}
	\end{center}
	\caption{Average ZDI magnetic field strength (top row) and field ratio (bottom row) against Rossby number for the magnetic dipole (left), quadrupole (center) and octupole (right) components. The subscript $\rm i=\{d,q,o\}$ represents each of the three components. Each point is colour coded by stellar mass. The range of solar values for each component is shown using a magenta strut. A three parameter fit (equation (\ref{eq:ActRot})) is performed for each component in the top row (solid red line). The best fit values can be found in table \ref{tab:FitParams}.}
	\label{fig:MagProps}
\end{figure*}

\section{The stellar sample and its magnetic properties}
\label{sec:MagProps}
The sample of stars we will use in this study is the one presented by \citet{See2019}. This sample consists of 85 stars that have had their magnetic fields mapped with ZDI. Some of these stars have been mapped at multiple epochs resulting in 151 magnetic maps. This collection of ZDI maps is drawn from many sources and represents nearly two decades of observations. The parameters (mass, radius, luminosity and rotation period) used for this work are listed in table 1 of \citet{See2019} and were generally taken from the original paper that the ZDI map was published in. Additionally, convective turnover times, $\tau_{\rm cz}$, were calculated using the formulation presented by \citet{Cranmer2011}\footnote{As noted in \citet{See2019}, \citet{Cranmer2011} state that this method of calculating convective turnover times is valid for stars with effective temperatures in the range 3300 K $\lesssim T_{\rm eff} \lesssim$ 7000 K. Although a number of our stars have $T_{\rm eff}< 3300$ K, they all lie in the saturated regime where the magnetic properties of stars, such as field strength, do not change significantly as a function of Rossby number. Therefore, this method of calculating convective turnover times will not greatly affect our results.}.

In addition to the parameters listed in table 1 of \citet{See2019}, we also require field strength values for the dipole, quadrupole and octupole components of the magnetic field for this work, i.e. $B_{\rm d}$, $B_{\rm q}$ and $B_{\rm o}$ in equation (\ref{eq:BrakingLaw}). In the MHD simulations of \citet{Finley2018}, $B_{\rm d}$, $B_{\rm q}$ and $B_{\rm o}$ correspond to field strengths at the rotation pole (or, equivalently in their simulations, the magnetic pole). However, these authors only considered axisymmetric field topologies whereas ZDI maps contain both axisymmetric and non-axisymmetric modes. Rather than using polar field values, we will instead use the stellar surface averaged unsigned field strength for each spherical harmonic mode\footnote{More formally, for each harmonic mode, we consider the power in the poloidal component of the component we are interested in. For example, using the formalism shown in appendix B of \citet{Folsom2018}, $\langle B_{\rm ZDI,q}\rangle$ would be calculated using the $\alpha_{\ell,m}$ and $\beta_{\ell,m}$ coefficients with $\ell=2$ \&  $m=\{0,1,2\}$ and all other coefficients set to zero.}. We shall denote these as $\langle B_{\rm ZDI,d}\rangle$, $\langle B_{\rm ZDI,q}\rangle$ and $\langle B_{\rm ZDI,o}\rangle$. Values for $\langle B_{\rm ZDI,d}\rangle$, $\langle B_{\rm ZDI,q}\rangle$ and $\langle B_{\rm ZDI,o}\rangle$ are shown in table \ref{tab:Values} for our entire sample along with citations to the original paper each ZDI maps is published in. 

Numerous studies have shown that magnetic activity indicators scale with the ratio of the rotation period over the convective turnover time, which is known as the Rossby number, ${\rm Ro} = P_{\rm rot}/\tau_{\rm cz}$. In the top row of fig. \ref{fig:MagProps}, we plot $\langle B_{\rm ZDI,d}\rangle$, $\langle B_{\rm ZDI,q}\rangle$ and $\langle B_{\rm ZDI,o}\rangle$ against $\rm Ro$. Each component follows a relatively tight power law relation at large Rossby numbers and appears to saturate at small Rossby numbers. This separation into saturated and unsaturated regimes is well known from X-ray studies \citep{Pizzolato2003,Wright2011,Wright2016,Stelzer2016,Wright2018}. We have also plotted solar values in each of these panels. These values are calculated using the Solar magnetograms from \citet{Vidotto2018} that cover most of cycle 24. For each magnetogram, we calculate a surface averaged poloidal dipole, quadrupole and octupole field strength. These are then averaged to determine the average poloidal dipole, quadrupole and octupole field strength over cycle 24 and plotted with magenta squares. The range of possible values over cycle 24 for each of these quantities is plotted with a magenta strut.

For each field component, we perform a three parameter fit of the form

\begin{equation}
\begin{split}
	\langle B_{\rm ZDI,i}\rangle & = B_{\rm sat,i} \hspace{25mm} {\rm for\ Ro<Ro_{crit,i}}  \\
	\langle B_{\rm ZDI,i}\rangle & = B_{\rm sat,i} \left(\frac{\rm Ro}{\rm Ro_{crit,i}}\right)^{\beta_{\rm i}} \hspace{5mm} {\rm for\ Ro\geq Ro_{crit,i}} 	
\end{split}	
\label{eq:ActRot}
\end{equation}
\begin{figure*}
	\begin{center}
	\includegraphics[trim=1cm 1cm 1cm 0cm,width=\textwidth]{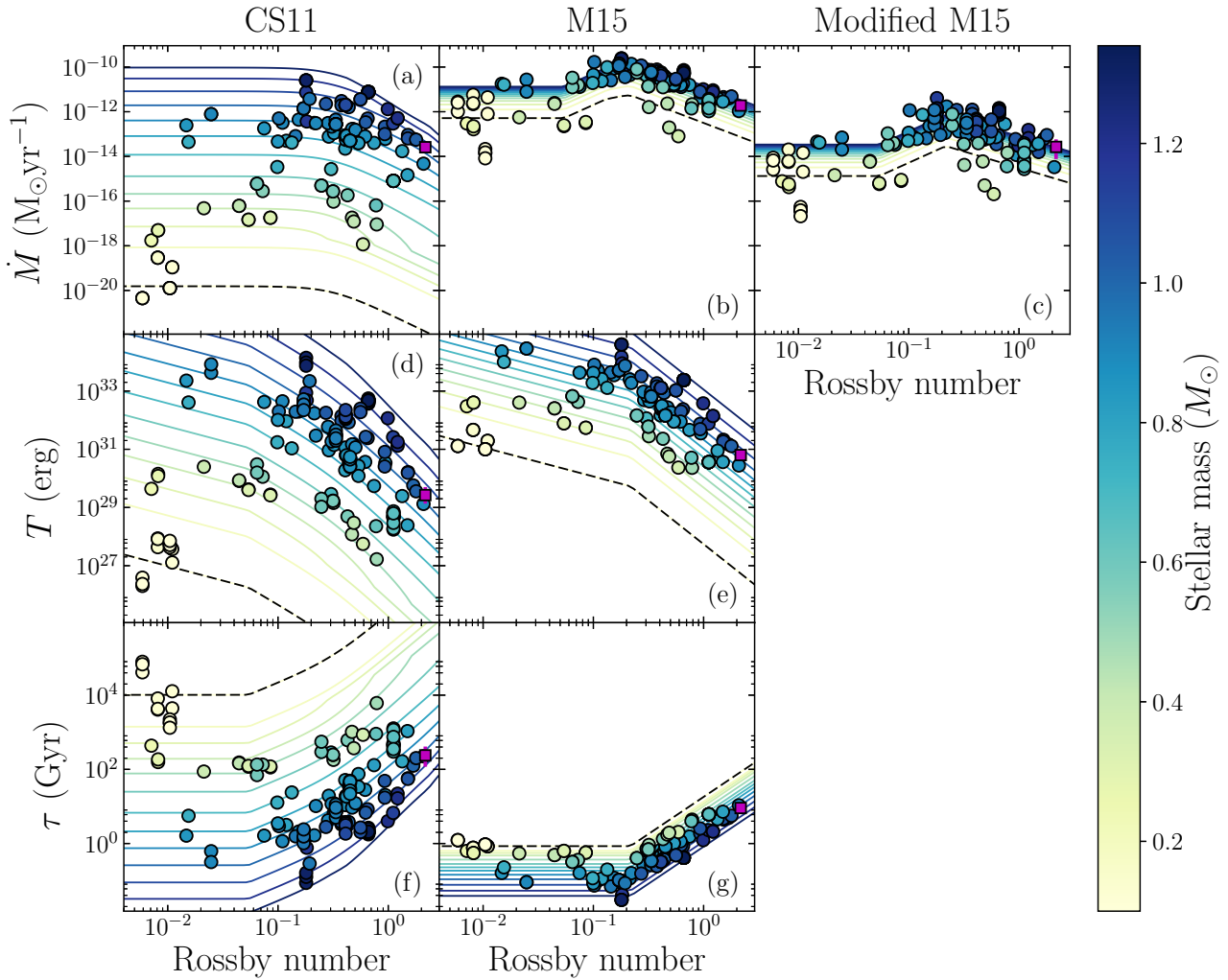}
	\end{center}
	\caption{Mass-loss rate (top), angular momentum-loss rate (middle) and spin-down time-scale (bottom) as a function of Rossby number. These quantities are calculated using (or are associated with) the CS11 (left), M15 (center) and modified M15 (right) models. Each point is colour coded by stellar mass. Solid lines correspond to the parameter of interest in each panel calculated for a given stellar mass over a range of Rossby numbers (see text for details). The solid lines are plotted in intervals of $\rm 0.1M_{\odot}$ from $\rm 0.1M_{\odot}$ to $1.3M_{\odot}$ and are also colour coded by stellar mass. The lines corresponding to $0.1M_{\odot}$ have been plotted with a dashed black line for visibility. The average solar value in each panel is shown with a magenta square and the range of variability is shown with a magenta strut when applicable.}
	\label{fig:LargePlot}
\end{figure*}
\begin{table}
\begin{minipage}{88mm}
	\begin{center}	
	\caption{Best fit parameters for equation (\ref{eq:ActRot})}
	\label{tab:FitParams}
	\begin{tabular}{lccc}
	\hline
	& $B_{\rm sat,i}$ & ${\rm Ro_{crit,i}}$ & $\beta_{\rm i}$\\
	\hline
	Dipole & 137$\pm 48$ & 0.05$\pm 0.02$ & -1.31$\pm 0.10$ \\
	Quadrupole & 73$\pm 21$ & 0.05$\pm 0.01$ & -1.25$\pm 0.08$ \\
	Octupole & 65$\pm 17$ & 0.05$\pm 0.01$ & -1.37$\pm 0.09$ \\
	\hline
\end{tabular}
\end{center}
\end{minipage}
\end{table}
where the subscript $\rm i = \{d,q,o\}$ represents each of the three components, $B_{\rm sat,i}$ is the field strength in the saturated regime, $\rm Ro_{crit,i}$ is the critical Rossby number separating the saturated and unsaturated regimes and $\beta_{\rm i}$ is the power law index in the unsaturated regime. As discussed in \citet{See2019}, we have excluded the stars with $\rm Ro \lesssim 0.012$ from these fits since they appear to display bimodal magnetic fields \citep{Donati2008,Morin2008,Morin2010} for which there is currently no definitive explanation. The best fit values can be found in table \ref{tab:FitParams} and the corresponding best fit curves are plotted in red in fig. \ref{fig:MagProps}. All three fits have similar $\rm Ro_{crit}$ and $\beta_{\rm}$ values within the error bars. The most notable difference in the fits is that the dipole component has a larger saturation field strength compared to the other two components. The saturation field strength is likely to be the least well constrained parameter however, since there are far fewer stars here compared to the unsaturated regime.  Additionally, due to observational biases, the Rossby number is correlated to stellar mass in our sample of stars. Although we have parameterised our fit in terms of Rossby number, there may be an additional dependence on stellar mass that is hard to disentangle from the dependence on Rossby number.

As well as the raw field strengths, we can also consider the field ratios defined in section \ref{sec:torque} for our sample of stars, i.e. $\mathcal{R}_{\rm i} = \langle B_{\rm ZDI,i}\rangle/\sum_{\rm j}\langle B_{\rm ZDI,j}\rangle$. These are plotted against Rossby number in the bottom row of fig \ref{fig:MagProps}. Each component shows a large amount of scatter with no obvious structure present. On average, $\mathcal{R}_{\rm d}$ has a higher value than $\mathcal{R}_{\rm q}$ or $\mathcal{R}_{\rm o}$ although this is not necessarily true for any individual star.

\section{Estimating mass-loss rates}
\label{sec:Mdots}
In this section, we estimate mass-loss rates for our sample of stars. Since estimating mass-loss rates for low-mass stars is difficult and model dependent, we explore two different methods. Although the main purpose of this section is to estimate mass-loss rates to compare to critical mass-loss rates in section \ref{sec:CriticalMdots}, we will also present the torques and spin-down time-scales associated with these methods since these are simple to calculate once mass-loss rates have been estimated. 

\subsection{Cranmer \& Saar (2011) method}
\label{subsec:CS11Mdots}
Our first method of estimating mass-loss rates uses the 1 dimensional model of \citet[][henceforth CS11]{Cranmer2011} which takes the stellar mass, radius, luminosity, rotation period and metallicity as inputs. We have chosen to use solar metallicity for all our stars for simplicity. This model estimates the energy generated from turbulent convective motions in the stellar interior. It then tracks this energy as it travels upwards  through the photosphere in the form of MHD waves. Eventually the energy is deposited along open field lines, heating up the local plasma and driving a hot coronal wind. 

Fig. \ref{fig:LargePlot}a shows the mass-loss rates of our sample estimated with the CS11 model. We have also plotted curves where each line corresponds to mass-loss rates for a fixed stellar mass over a range of Rossby numbers (similar curves are plotted on the other panels of fig. \ref{fig:LargePlot}). These lines are included to illustrate the behaviour of the CS11 model over a range of fixed masses and are intended to aid the reader in interpreting the data points. Additionally, they allow for a rough estimate of mass-loss rates in regions of parameter space that our ZDI sample does not cover. For each line, stellar radii and luminosities, which are required by the CS11 model, are estimated by interpolating over the grid of stellar evolution models of \citet{Baraffe2015} at an age of 2 Gyr. This age is chosen to be older than the zero age main sequence but younger than the main sequence turnoff for the types of stars in our sample. We have therefore assumed that the stellar radius and luminosity are only a function of mass, and not age or rotation, when calculating these curves which is approximately true on the main sequence. This method produces radius and luminosity estimates that are broadly representative of the stars we are interested in. Although it is a relatively simple method of estimating radii and luminosities, it is appropriate since the curves are simply illustrative and included only to help the reader interpret the data points. Overall, the data points in fig. \ref{fig:LargePlot} follow the trends shown by the lines although there will be small deviations due to a number of different factors, e.g. the fact that the stars in our sample have a range of ages. For a given stellar mass, we see that the mass-loss rates follow the activity-rotation relation shape described in section \ref{sec:MagProps}. The most striking feature is the range of predicted mass-loss rates, spanning around ten orders of magnitude. The main determinant of the mass-loss rate for the CS11 model is the stellar mass, with rotation (or Rossby number) having a secondary effect.

Using these mass-loss rates, we can calculate torques, $T_{\rm CS11}$, and spin-down time-scales, $\tau_{\rm CS11}$. $T_{\rm CS11}$ is calculated using the F18 braking law (equation (\ref{eq:BrakingLaw})) and are shown in fig. \ref{fig:LargePlot}d. $\tau_{\rm CS11}$ is given by $I_{\star}\Omega_{\star}/T_{\rm CS11}$, where $I_{\star}$ is the moment of inertia of the star. Again, we use the stellar evolution models of \citet{Baraffe2015} to obtain moments of inertia for our sample of stars. $\tau_{\rm CS11}$ is plotted in fig. \ref{fig:LargePlot}f. Similarly to the mass-loss rate, we have plotted $T_{\rm CS11}$ and $\tau_{\rm CS11}$ curves in panels (d) and (f). When calculating the $T_{\rm CS11}$ curves, we used the fits to our magnetic field data (equation (\ref{eq:ActRot})) to determine the magnetic properties required in equation (\ref{eq:BrakingLaw}). Due to the low mass-loss rates estimated by this model and the correspondingly low torques, the characteristic spin-down time-scales are large, especially at the lowest masses. Given that M dwarfs are known to spin down on time-scales shorter than those shown in fig. \ref{fig:LargePlot}f \citep[e.g.][]{Douglas2017}, one might interpret the large $\tau_{\rm CS11}$ values for M dwarfs as evidence that the CS11 model substantially underestimates the mass-loss rates for low-mass stars.

\subsection{Rotation evolution method}
\label{subsec:RotEvoMdots}
Our second method is to determine the mass-loss rate required by the F18 braking law in order to reproduce the rotation period evolution seen in open clusters. We will base this on the rotation period evolution model of \citet[][henceforth M15]{Matt2015} which was tuned to broadly reproduce observed rotation period distributions in open clusters, e.g. see their fig. 2. As discussed in section \ref{sec:MagProps}, many forms of magnetic activity can be parameterised in terms of the Rossby number but with different scalings depending on whether the star is in the so-called saturated or unsaturated regime. Motivated by this idea, \citet{Matt2015} assumed that magnetic terms in their spin-down torque can also be parameterised by Rossby number with different scalings in the saturated and unsaturated regimes (see their equations (4) and (5)). The resulting torque has the form

\begin{equation}
\begin{split}
	T & = 100\ T_0 \left(\frac{\Omega_{\star}}{\Omega_{\odot}}\right) \hspace{19mm} {\rm for\ Ro< Ro_{crit}} \\
	T & = T_0 \left(\frac{\tau_{\rm cz}}{\tau_{\rm cz\odot}}\right)^2 \left(\frac{\Omega_{\star}}{\Omega_{\odot}}\right)^{3} \hspace{9mm} {\rm for\ Ro\geq Ro_{crit}}.
\end{split}
\label{eq:Matt15Torque}
\end{equation}
Here, $T_0$ is an additional mass (and radius) dependent factor that is given by

\begin{equation}
	T_0 = 6.3\times 10^{30}\ {\rm erg} \left(\frac{r_{\star}}{r_{\odot}}\right)^{3.1} \left(\frac{M_{\star}}{M_{\odot}}\right)^{0.5}.
	\label{eq:T0}
\end{equation}
This additional mass dependence is required to explain observations that show that the lowest mass stars take much longer to spin down compared to their higher mass counterparts. M15 chose their critical Rossby number to be $\rm Ro_{crit} = 0.2$. It is worth noting that this value is larger than the $\rm Ro_{crit}$ values we obtained in our three parameter fits in section \ref{sec:MagProps}. For the Sun, this model estimates a spin-down torque of $6.3\times 10^{30} {\rm erg}$ which is the value one finds if one assumes the Sun is a solid body and spinning down according to \citet{Skumanich1972}, i.e. $\Omega \propto t^{-1/2}$ \citep[see section 4.3 of][ for further details]{Finley2018}. This value for the solar spin-down torque is similar to the values used in other rotation evolution models \citep[see fig. 1 of ][for a comparison]{Matt2015}.

In fig. \ref{fig:LargePlot}e, we plot the torque for our sample using equation (\ref{eq:Matt15Torque}). These torques are larger than those estimated using the CS11 model (fig. \ref{fig:LargePlot}d) with the largest disagreement occurring for the lowest mass stars. In fig. \ref{fig:LargePlot}b, we plot the mass-loss rate required by the F18 braking law to reproduce the torque used by M15, i.e. the mass-loss rate at which equations (\ref{eq:Torque}) \& (\ref{eq:BrakingLaw}) equals equation (\ref{eq:Matt15Torque}). These mass-loss rates are much higher than those estimated using the CS11 model. This is unsurprising given the lower torques estimated when using the CS11 mass-loss rates. Lastly, we plot the spin-down time-scale associated with this model, $\tau_{\rm M15}$, in fig. \ref{fig:LargePlot}g. The $\tau_{\rm M15}$ values are generally smaller than the $\tau_{\rm CS11}$ values due to the larger $T_{\rm CS11}$ values.

\begin{figure}
	\begin{center}
	\includegraphics[trim=1cm 1cm 1cm 0cm,width=\columnwidth]{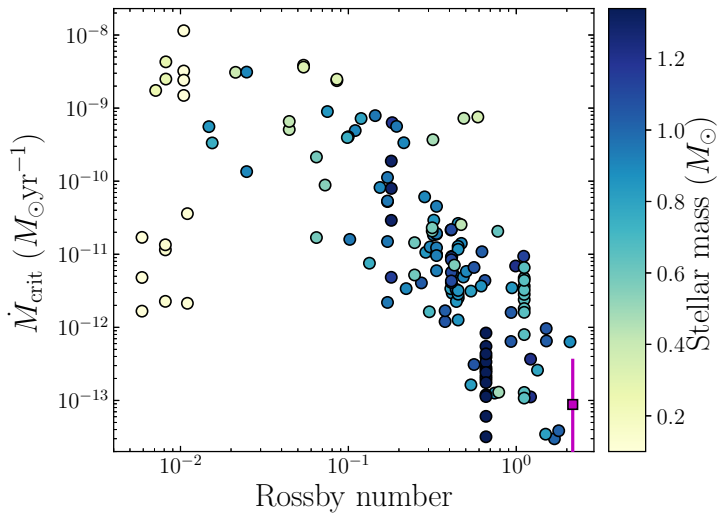}
	\end{center}
	\caption{Critical mass-loss rate (equation (\ref{eq:MDotCrit})) against Rossby number. Points are coloured coded by stellar mass. The average solar $\dot{M}_{\rm crit}$ is shown with a magenta square and the range over cycle 24 is shown by the magenta bar. The lower limit of this range extends off the plot and has a value of $\dot{M}_{\rm crit}=5.7\times 10^{-16} M_{\odot}\ {\rm yr^{-1}}$ (the full range and variability of the magenta bar is shown by the magenta curve in fig. \ref{fig:SolarPlot}). For each star, the angular momentum-loss rate is dominated by the dipole component of stellar magnetic field if its actual mass-loss rate is below the critical mass-loss rate shown here.}
	\label{fig:MDotCrit}
\end{figure}

\subsection{Modified rotation evolution method}
\label{subsec:ModRotEvoMdots}
There is a striking problem with the model presented in section \ref{subsec:RotEvoMdots}. If one calculates the solar spin-down torque using the F18 braking law and the observed solar mass-loss rate, one obtains a value of $T={\rm 2.5\times 10^{29} erg}$ which is a factor of 25 smaller than the value predicted by the rotation evolution model of \citet{Matt2015}. Conversely, if one calculates the mass-loss rate for the Sun in the manner described in section \ref{subsec:RotEvoMdots}, using the parameters from table \ref{tab:Sun}, one obtains a value of $\dot{M}=1.8\times 10^{-12} M_{\odot}\ {\rm yr}^{-1}$. This is a factor of $\sim 70$ bigger than the observed solar mass-loss rate. Additionally, this method also estimates mass-loss rates that are much larger than mass-loss rates inferred from Ly-$\alpha$ observations (see \citep{Vidotto2016} for a sample of stars that have both ZDI maps and mass-loss rates inferred from Ly-$\alpha$ observations). Clearly, there is a discrepancy between the F18 braking law and the rotation evolution models \citep[also see discussion in ][]{Finley2018Sun,Finley2019}.

\begin{figure*}
	\begin{center}
	\includegraphics[trim=1cm 2cm 1cm 0cm,width=\textwidth]{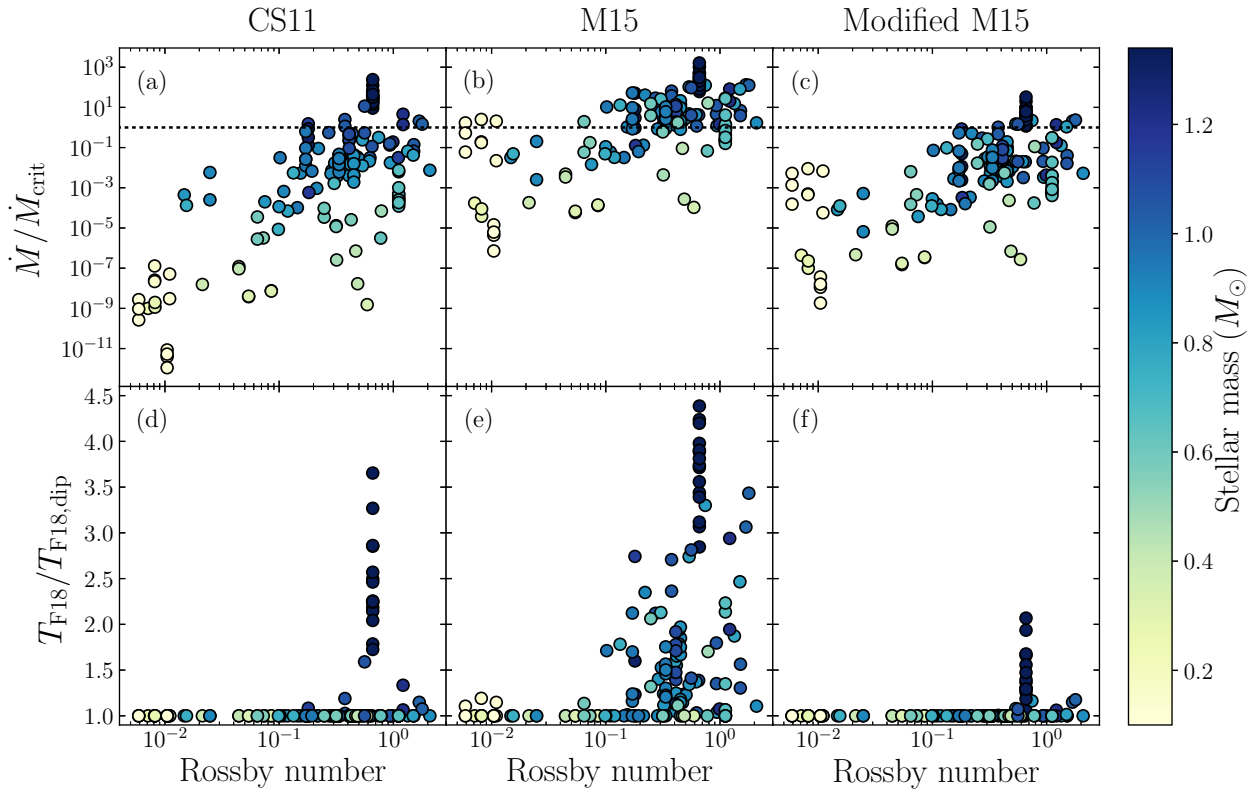}
	\end{center}
	\caption{Top: the ratio of mass-loss rate to critical mass-loss rate against Rossby number. Points above the dotted $\dot{M} = \dot{M}_{\rm crit}$ line have a non-dipolar field contribution to spin-down torque. Bottom: The ratio of the torque calculated using the F18 braking law to the torque calculated using just the dipole component of the F18 braking law against Rossby number. Both quantities are calculated using the CS11 (left), rotation evolution (middle) and modified rotation evolution (right) methods. Points are colour coded by stellar mass.}
	\label{fig:Compare}
\end{figure*}

Although the origin of this discrepancy is unclear, one possible explanation is that the MHD models used by F18 may under predict the torques. One solution is to include a multiplicative correction factor in the F18 braking law.  In fig. \ref{fig:LargePlot}c, we recalculate mass-loss rates using the method from section \ref{subsec:RotEvoMdots} but including a multiplicative factor of 25 in the F18 braking law. This value is chosen such that the solar mass-loss rate is recovered when using the method described in section \ref{subsec:RotEvoMdots} for the Sun. Including the multiplicative factor reduces the estimated mass-loss rates by a factor of between 50 and 390, the value of which depends on the values of $\langle B_{\rm ZDI,d}\rangle$, $\langle B_{\rm ZDI,q}\rangle$ \& $\langle B_{\rm ZDI,o}\rangle$ for each star. We note that if the F18 braking law does underestimate spin-down torques by a factor of 25, then the torques in fig. \ref{fig:LargePlot}d should all be larger by this amount.

\section{Critical mass-loss rates}
\label{sec:CriticalMdots}
Having estimated mass-loss rates in section \ref{sec:Mdots}, we can now compare them to the critical mass-loss rates as defined in section \ref{sec:torque}. Fig. \ref{fig:MDotCrit} shows that the overall trend is for the critical mass-loss rate to decrease as a function of Rossby number. This decrease can be attributed to the dependence of $\dot{M}_{\rm crit}$ on $B_{\star}^2$ (see equation (\ref{eq:MDotCrit})). Physically, this is because stars with strong magnetic fields require a correspondingly large mass-loss rate for the Alfv\'{e}n radius to be small enough for non-dipolar fields to contribute to the spin-down torque. There is also a large amount of scatter in fig. \ref{fig:MDotCrit} that can be attributed to the scatter in $B_{\star}^2$, $\mathcal{R}_{\rm d}$ and $\mathcal{R}_{\rm q}$. Lastly, there is a departure from the overall trend at the lowest Rossby numbers that is caused by the bimodal magnetic fields of the lowest mass M-dwarfs \citep{Donati2008,Morin2008,Morin2010}.

In the top row of  fig. \ref{fig:Compare}, we show the ratio of the mass-loss rate to the critical mass-loss rate against Rossby number for each of the mass-loss rate estimates outlined in section \ref{sec:Mdots}. The dotted lines indicate $\dot{M} = \dot{M}_{\rm crit}$. For the CS11 model, we find that the majority of the stars have $\dot{M}<\dot{M}_{\rm crit}$ and, consequently, only the dipole component of the field is required to properly estimate the torque using the F18 braking law for most stars using this method. The few stars that do have $\dot{M}>\dot{M}_{\rm crit}$ are the highest mass stars since the mass-loss rates estimated by CS11 model has a strong dependence on stellar mass. It should be noted that a significant number of the points in the $\dot{M}>\dot{M}_{\rm crit}$ regime are for one star, $\tau$ Boo. This is the star that we have the most ZDI maps for and is also the highest mass star in our sample. In contrast to the CS11 model, a majority of the stars have $\dot{M}>\dot{M}_{\rm crit}$ when using the rotation evolution method (M15) to estimate mass-loss rates. However, as discussed in section \ref{subsec:ModRotEvoMdots}, these mass-loss rates are likely to be too high. Using the modified rotation evolution method to estimate mass-loss rates results in a significant reduction in the number of stars that have $\dot{M}>\dot{M}_{\rm crit}$.

Fig. \ref{fig:Compare} demonstrates that there may be regimes where the mass-loss rates of low-mass stars are sufficiently high that non-dipolar field modes need to be accounted for to properly calculate their spin-down torques. It will be interesting to determine how different the spin-down torques for these stars are if we only accounted for their dipole fields. In the bottom row of fig. \ref{fig:Compare}, we plot the ratio of the torque calculated using the full F18 braking law, $T_{\rm F18}$, to the torque calculated using just the dipole component, {$T_{\rm F18,dip}$, i.e. equation \ref{eq:BrakingLawDip}. This ratio is calculated using the mass-loss rate estimates from section \ref{sec:Mdots}, i.e. the mass-loss rates shown in figs. \ref{fig:LargePlot}a, \ref{fig:LargePlot}b \& \ref{fig:LargePlot}c. By definition, this ratio is equal to 1 when $\dot{M}<\dot{M}_{\rm crit}$. However, when $\dot{M}>\dot{M}_{\rm crit}$, $T_{\rm F18}$ is bigger than $T_{\rm F18,dip}$. This can be seen in fig. \ref{fig:SolarBrakingLaw} where the dipole only line (red dot dashed line) drops below the upper envelope of the three curves for $\dot{M}>\dot{M}_{\rm crit}$. Physically, the reason that $T_{\rm F18}>T_{\rm F18,dip}$ is because the Alfv\'{e}n radii is sufficiently small, at high mass-loss rates, for the non-dipolar fields to contribute to the spin-down torque. However, since $T_{\rm F18,dip}$ does not account for the flux in non-dipolar modes, the resulting torque is smaller. $T_{\rm F18}/T_{\rm F18,dip}$ can reach values of around 3.7, 4.4 and 2.1 respectively for the CS11, M15 and modified M15 methods. However, this is skewed by $\tau$ Boo. If we put $\tau$ Boo aside, $T_{\rm F18}/T_{\rm F18,dip}$ never exceeds 1.6 for the CS11 and modified M15 methods.

It is interesting to note that, generally, the non-dipolar field modes only become important at large Rossby numbers. For the CS11 method, this is mainly an effect of the stellar mass. As already noted, the stars with the highest mass-loss rates, i.e. those most likely to have $\dot{M}>\dot{M}_{\rm crit}$, are the highest mass stars. In our sample, these happen to be the ones with the largest Rossby numbers. For the rotation evolution and modified rotation evolution models, the explanation is slightly different. We have already noted that the critical mass-loss rate has a dependence on $B_{\star}^2$. For these models, both $\dot{M}$ (for a given mass) and $B_{\star}^2$ decrease as a function of Rossby number in the unsaturated regime. However, $B_{\star}^2$, and hence $\dot{M}_{\rm crit}$, decreases more steeply than $\dot{M}$. As such, broadly speaking, $\dot{M}$ becomes larger than $\dot{M}_{\rm crit}$ as Rossby number increases. This also predicts that Alfv\'{e}n radii decrease as stars spin down.

\section{Uncertainties}
\label{sec:Uncertainties}
In this section, we discuss the uncertainties, caveats and open questions relating to the methods and models we have used in this paper as well as where future progress may impact our conclusions.

\begin{figure*}
	\begin{center}
	\includegraphics[trim=0cm 1cm 0cm 0.5cm,width=0.75\textwidth]{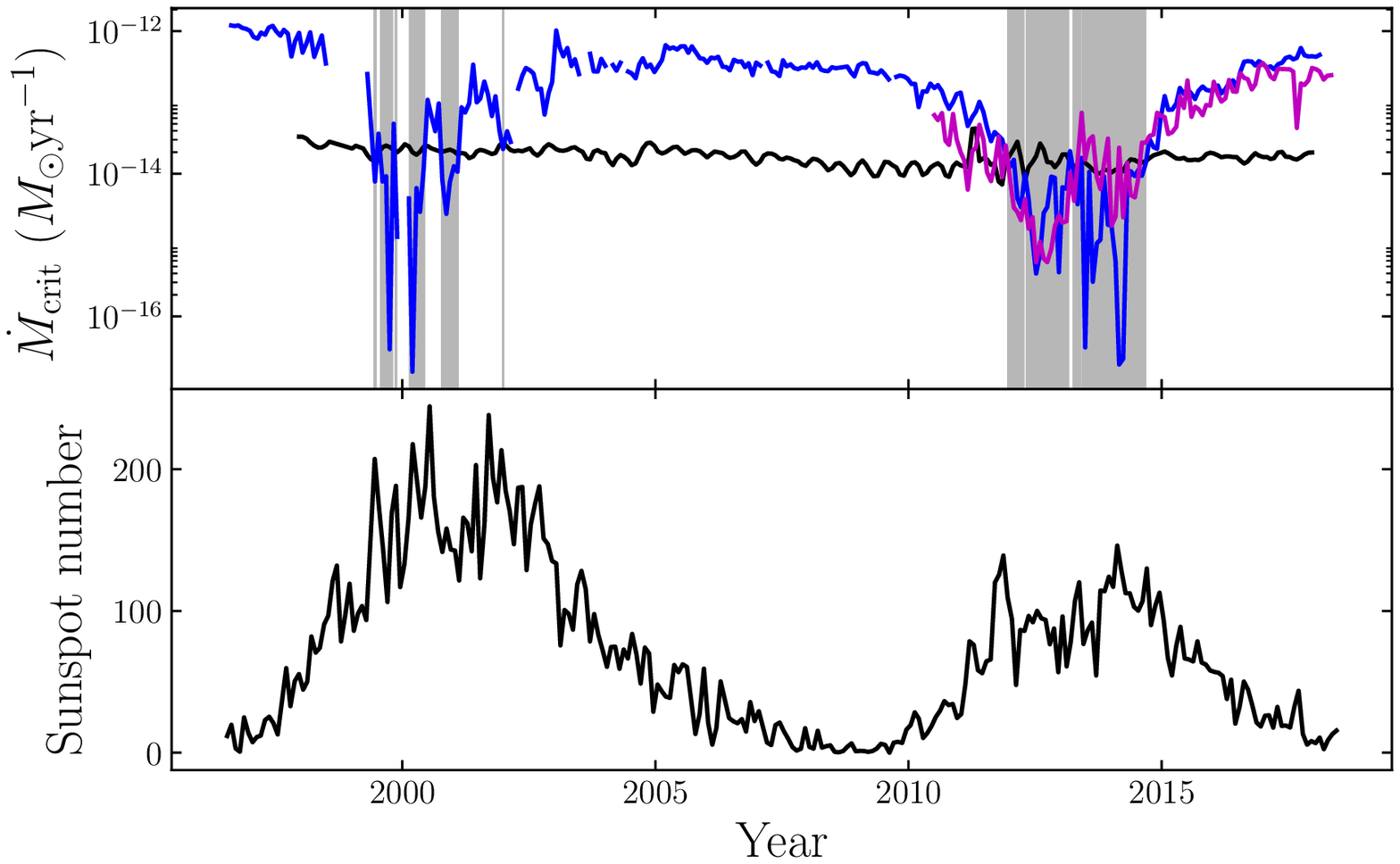}
	\end{center}
	\caption{Top: critical mass-loss rate for the Sun calculated using magnetic data from \citet[][magenta line]{Vidotto2018} and \citet[][blue line]{Finley2018Sun}. The solar mass-loss rate estimated by \citet{Finley2018} is also shown (black line). Regions where the mass-loss rate is greater than the critical mass-loss rate are shaded in gray. Bottom: Sunspot number.}
	\label{fig:SolarPlot}
\end{figure*}

\subsection{Zeeman-Doppler imaging}
\label{subsec:ZDI}
Modern ZDI codes express magnetic fields in terms of a spherical harmonic decomposition (e.g. see appendix B of \citet{Folsom2018}). It is well known that ZDI does not recover the magnetic field associated with small scale structures, e.g. starspots, due to flux cancellation effects \citep{Reiners2009,Morin2010,See2019}. Typically, ZDI maps have a maximum spherical harmonic degree of $\ell_{\rm max}=5$ to 15. \citet{Lehmann2018ZDI} recently showed that even the field strengths associated with low order $\ell$ modes, including the dipole, quadrupole and octupole modes, can be systematically underestimated by factors of a few by ZDI. The possibility of underestimated field strengths has two main consequences for our results.

The first consequence is that the spin-down torque will be underestimated due to the dependence of the F18 braking law on magnetic field strength. Underestimated field strengths may partially alleviate a problem noted by \citet{Finley2019}. Similarly to this work, these authors estimated the spin-down torques for a number of stars with ZDI maps using the F18 braking law. However, they used mass-loss rates inferred from Ly-$\alpha$ observations (see e.g. \citet{Wood2014}). The torques they estimated using the F18 braking law were smaller than the torques estimated by the M15 rotation evolution model by a factor of $\sim3-30$. Closer agreement between these two torque estimates could be achieved if the dipolar, quadrupolar and octupolar field strengths are larger than those inferred by ZDI. Their results are also consistent with our results from section \ref{subsec:RotEvoMdots}. The mass-loss rates we estimate using the rotation evolution method are much larger than those inferred from  Ly-$\alpha$. As with the results of \citet{Finley2019}, the discrepancy between rotation evolution based estimates of stellar torques and braking law based estimates can be partially alleviated by larger field strengths. However, it is unlikely that the discrepancy can be fully accounted for by underestimated field strengths from ZDI.

The second consequence relates to the fact that $\dot{M}_{\rm crit} \propto B_{\star}^2$. If the true field strengths are a factor of several higher than that recovered by ZDI, the critical mass-loss rate would increase by roughly an order of magnitude. Consequently, even fewer stars than suggested in section \ref{sec:CriticalMdots} will have $\dot{M}>\dot{M}_{\rm crit}$. 

\subsection{Mass-loss rate estimates}
\label{subsec:MDotModels}
Estimating mass-loss rates is notoriously difficult because the mechanisms that drive stellar winds are still poorly understood. Although the CS11 model is one of the more sophisticated models currently available for estimating mass-loss rates, it still has limitations. It is based on many previous observations and theoretical works and uncertainty in those works will propagate through into the final mass-loss rate estimate. For example, one key part of the model that remains relatively unconstrained is the magnetic characteristics of low-mass stars. Parameters such as the amount of energy flux in Alfv\'en waves, the fraction of the stellar surface covered in open flux tubes and the rate at which these flux tubes expand above the stellar surface all play an important role in determining the overall mass-loss rate but remain uncertain. Other details such as the terminal wind speed or the location of the transition region are also hard to determine reliably and will contribute to the overall uncertainty of this model. We refer the interested reader to section 6 of CS11 for a much more comprehensive discussion of these uncertainties.

Our second method of estimating mass-loss rates relies on rotation period evolution models. Although models are now able to capture the overall rotation evolution behaviour of low-mass stars, none can yet fit all the available data. Additionally, the rotation evolution models are also not well constrained in areas of parameter space where rotation period data is sparse, particularly for the lowest mass and the oldest stars. In the model of \citet{Matt2015}, the rotation periods of the lowest mass stars are not reproduced well which can be seen when comparing the model to observations from the Hyades \citep[see fig. 14 in][]{Douglas2016} and Praesepe \citep[see fig. 11 in][]{Douglas2017}. Outside of the lowest mass M dwarfs, the disagreement between the M15 model and observed rotation periods can be up to $\sim$50\% in some mass and age ranges. Our mass-loss rate estimates that are based on the rotation evolution models (section \ref{subsec:ModRotEvoMdots}) will therefore be correspondingly uncertain. However, since the model reproduces the broad features seen in rotation period distributions with a relatively simple torque prescription, it is ideal for the purposes of this work.

\subsection{Stellar variability}
\label{subsec:Variability}
In this work, we have estimated mass-loss and angular momentum-loss rates at single instances in time (or at a few instances for stars with multiple ZDI maps). However, stellar magnetic activity is known to be time variable. Therefore, their mass-loss and angular momentum-loss rates are also time variable. For example, the angular momentum-loss rate of the Sun is expected to vary over a range of time-scales \citep{Pinto2011,Reville2017,Finley2018Sun,Perri2018}. As noted in section \ref{sec:torque}, the Sun has a mass-loss rate that is smaller than its critical mass-loss rate. However, that was calculated using averages of the solar dipolar, quadrupolar and octupolar fields over $\sim$8 years using the data from \citet{Vidotto2018}. In the top panel of fig. \ref{fig:SolarPlot}, we show the critical mass-loss rate as a function of time for the Sun. This is done using data from \citet{Vidotto2018} and \citet{Finley2018Sun} shown in magenta and blue respectively\footnote{The solar magnetograms of \citet{Vidotto2018} consist of the radial, meridional and azimuthal components while \citet{Finley2018Sun} only consider the radial component in their work. So far in this work, we have used the magnetograms of \citet{Vidotto2018} because they provide a better comparison to ZDI maps that also contain all 3 vector components. However, the dataset studied by \citet{Finley2018Sun} covers a longer time period and so we have included both in fig. \ref{fig:SolarPlot}.}. From 2010 to 2018, where data is available from both studies, the two critical mass-loss rate estimates are broadly comparable. We also plot the solar mass-loss rate determined from in situ spacecraft measurements \citep{Finley2018Sun} with a black line. Times when the mass-loss rate exceeds the critical mass-loss rate (estimated from the \citet{Finley2018Sun} data) are shaded in gray. For reference, the bottom panel shows sunspot number indicating periods of maximum activity around 2001 and 2013. We see that the solar mass-loss rate does not exceed the critical mass-loss rate for the majority of the last two solar cycles suggesting that the dipole magnetic field dominates the solar torque most of the time. However, the non-dipolar components become important around solar maximum. This is due to the growing quadrupolar field and shrinking dipolar field at solar maximum \citep{DeRosa2012}.

The magnetic fields of stars that have been mapped with ZDI over multiple epochs are also known to evolve over time \citep[e.g.][]{Jeffers2014,Saikia2016,Lavail2018,Saikia2018}. Correspondingly, their estimated torques also change as a function of time. As with the Sun, this means other stars can spend some times with $\dot{M}>\dot{M}_{\rm crit}$ and other times with $\dot{M}<\dot{M}_{\rm crit}$. In the context of long-term rotation period evolution, models are only sensitive to the spin-down torque averaged over time-scales of Myr or more so comparisons with instantaneous torque estimates will always be uncertain.

\subsection{Braking law}
\label{subsec:BrakingLaw}
While the F18 braking law allows for a rapid assessment of the spin-down torque of a star, there are still areas in which it can be improved. For instance, the MHD simulations on which this braking law is based assume a fixed coronal temperature and use the polytropic assumption which can have a small impact on the resulting wind solutions \citep{Pantolmos2017}. Additionally, F18 only investigated axisymmetric field configurations while we have also included non-axisymmetric field components in this work. By using the F18 braking law, we have effectively moved power from the non-axisymmetric modes to the axisymmetric ones. Although the effect of including non-axisymmetric modes into braking law studies like that of F18 is not entirely understood, we do not expect that it would drastically change our results \citep[also see section 5.1 of][]{Finley2018Sun}.

Lastly, using Ulysses data, \citet{Finley2018Sun} showed that the simulations used to construct the F18 braking law appear to underestimate the open flux for a given input solar magnetogram. This problem is not unique to the F18 simulations \citep[e.g.][]{Linker2017}. Correspondingly, the F18 spin-down torque estimates are also probably underestimated by a factor of a few just due to the open flux problem. The reason for this is not clear but any potential solution may increase the contribution of non-dipolar modes to the spin-down torque. In order to increase the open flux, the height of the last closed field loop must decrease (this radius is variously called the opening radius or source surface radius in the literature depending on the context). Since non-dipolar fields decay more rapidly than dipolar fields as a function of radius, the contribution of non-dipolar modes to the open flux is increased by opening up magnetic flux closer to the stellar surface. Unfortunately, it is currently difficult to quantify the magnitude of this effect but it should be kept in mind.

\section{Discussion and conclusion}
\label{sec:Conclusions}
Using a sample of stars that have been mapped with Zeeman-Doppler imaging and the braking law of \citet{Finley2018}, we have investigated whether non-dipolar magnetic fields contribute significantly to stellar spin-down.  In order to use this braking law, mass-loss rates have to be estimated for each of the stars. In general, large mass-loss rates are required for non-dipolar fields to contribute to stellar spin-down. We quantify this in terms of a critical mass-loss rate, $\dot{M}_{\rm crit}$, that depends on the field strength and geometry of the star. If a star has a mass-loss rate smaller than $\dot{M}_{\rm crit}$, only the dipolar field mode is required to calculate the spin-down torque. However, higher order field modes need to be accounted for when stars have mass-loss rates larger than $\dot{M}_{\rm crit}$. We used two methods to estimate mass-loss rate.

The first method uses the model of \citet{Cranmer2011}. Using these mass-loss rates, we find that the non-dipolar magnetic field modes do not contribute to spin-down for the majority of the stars in our sample. The stars that do have mass-loss rates larger than $\dot{M}_{\rm crit}$ are the highest mass stars in our sample. This is due to the strong dependence of mass-loss rate on stellar mass in the \citet{Cranmer2011} model. 

The second method estimates mass-loss rates by determining the mass-loss rates required by the \citet{Finley2018} braking law to reproduce the spin-down torques from the rotation evolution model of \citet{Matt2015}. This method produces much higher mass-loss rates than the model of \citet{Cranmer2011}. Consequently, a majority of the sample have mass-loss rates larger than $\dot{M}_{\rm crit}$. However, this method overestimates the solar mass-loss rate. This is likely because the \citet{Finley2018} braking law is under-predicting the spin-down torque. To reproduce the observed solar mass-loss rate using this method, a multiplicative factor of 25 needs to be included in the \citet{Finley2018} braking law. Once this is included, far fewer stars have mass-loss rates that exceed $\dot{M}_{\rm crit}$. In this model, the stars where high order field modes need to be accounted for are those at large Rossby number.

Our core conclusion is therefore that non-dipolar magnetic fields do not contribute significantly to stellar spin-down for the majority of low-mass stars. However, there are stars in some parameter regimes, whose mass-loss rates are estimated to be particularly large, for which this may not be true. This result is based on the mass-loss rate estimates of two models although one could in principle conduct this sort of study using any mass-loss rate model. Indeed, different methods of estimating mass-loss rates may predict different parameter regimes in which non-dipolar fields need to be accounted for. It is also worth noting that this conclusion assumes that the mass-loss rate of a star is known. In reality, small-scale fields are important for heating and determining the mass-loss rate \citep[e.g.][]{Cranmer2011,Suzuki2013}. However, for a given mass-loss rate, only the dipole field is important for stellar spin-down for most stars.

Lastly, it is worth discussing our results in the context of rotation evolution models. In recent years, higher magnetic field complexity has been invoked when a reduced torque is required by rotation evolution models to fit the observed period distributions in open clusters. The justification for this is that MHD simulations have shown that, all else being equal, stars with higher order spherical harmonic field modes have vastly reduced torques \citep{Reville2015,Garraffo2016}. For example, \citet{Garraffo2018} cite the higher field complexity of rapid rotators as evidence that they should also have reduced spin-down torques. However, the MHD simulations for computing torques conducted before those of \citet{Finley2017,Finley2018} have generally only considered a single spherical harmonic order in each simulation. In reality, the magnetic fields of stars are a superposition of many spherical harmonic modes. While it is true that some stars within our sample have more magnetic energy at higher order spherical harmonic modes, i.e. fields associated with smaller spatial scales, all of the stars contain a non-negligible dipole component (see fig 2a). This is the reason why non-dipolar modes do not significantly contribute to stellar spin down for the majority of our stars.

\acknowledgments
The authors would like to thank the anonymous referee for useful comments that helped improved the quality of this manuscript. Sunspot number data is from WDC-SILSO, Royal Observatory of Belgium, Brussels. VS, SPM and AJF acknowledge funding from the European Research Council (ERC) under the European Unions Horizon 2020 research and innovation programme (grant agreement No 682393 AWESoMeStars). SBS acknowledges funding via the Austrian Space Application Programme (ASAP) of the Austrian Research Promotion Agency (FFG) within ASAP11, the FWF NFN project S11601-N16 and the sub-project S11604-N16. SJ acknowledges the support of the  German Science Foundation (DFG) Research Unit FOR2544 ``Blue Planets around Red Stars'', project JE 701/3-1 and DFG priority program SPP 1992 "Exploring the Diversity of Extrasolar Planets (RE 1664/18). AAV  acknowledges funding received from the Irish Research Council Laureate Awards 2017/2018.

\bibliographystyle{yahapj}
\bibliography{ZDITorques}

\begin{thebibliography}{}
\providecommand\natexlab[1]{#1}
\providecommand\JournalTitle[1]{#1}

\bibitem[{{Amard} {et~al.}(2016){Amard}, {Palacios}, {Charbonnel}, {Gallet}, \&
  {Bouvier}}]{Amard2016}
{Amard}, L., {Palacios}, A., {Charbonnel}, C., {Gallet}, F., \& {Bouvier}, J.
  2016,
  \href{http://dx.doi.org/10.1051/0004-6361/201527349}{\JournalTitle{A\&A},
  587, A105}

\bibitem[{{Baraffe} {et~al.}(2015){Baraffe}, {Homeier}, {Allard}, \&
  {Chabrier}}]{Baraffe2015}
{Baraffe}, I., {Homeier}, D., {Allard}, F., \& {Chabrier}, G. 2015,
  \href{http://dx.doi.org/10.1051/0004-6361/201425481}{\JournalTitle{A\&A},
  577, A42}

\bibitem[{{Blackman} \& {Owen}(2016)}]{Blackman2016}
{Blackman}, E.~G., \& {Owen}, J.~E. 2016,
  \href{http://dx.doi.org/10.1093/mnras/stw369}{\JournalTitle{MNRAS}, 458,
  1548}

\bibitem[{{Boro Saikia} {et~al.}(2015){Boro Saikia}, {Jeffers}, {Petit},
  {Marsden}, {Morin}, \& {Folsom}}]{Saikia2015}
{Boro Saikia}, S., {Jeffers}, S.~V., {Petit}, P., {et~al.} 2015,
  \href{http://dx.doi.org/10.1051/0004-6361/201424096}{\JournalTitle{A\&A},
  573, A17}

\bibitem[{{Boro Saikia} {et~al.}(2016){Boro Saikia}, {Jeffers}, {Morin},
  {Petit}, {Folsom}, {Marsden}, {Donati}, {Cameron}, {Hall}, {Perdelwitz},
  {Reiners}, \& {Vidotto}}]{Saikia2016}
{Boro Saikia}, S., {Jeffers}, S.~V., {Morin}, J., {et~al.} 2016,
  \href{http://dx.doi.org/10.1051/0004-6361/201628262}{\JournalTitle{A\&A},
  594, A29}

\bibitem[{{Boro Saikia} {et~al.}(2018){Boro Saikia}, {Lueftinger}, {Jeffers},
  {Folsom}, {See}, {Petit}, {Marsden}, {Vidotto}, {Morin}, {Reiners}, {Guedel},
  \& {BCool Collaboration}}]{Saikia2018}
{Boro Saikia}, S., {Lueftinger}, T., {Jeffers}, S.~V., {et~al.} 2018,
  \href{http://dx.doi.org/10.1051/0004-6361/201834347}{\JournalTitle{A\&A},
  620, L11}

\bibitem[{{Brown} {et~al.}(1991){Brown}, {Donati}, {Rees}, \&
  {Semel}}]{Brown1991}
{Brown}, S.~F., {Donati}, J.-F., {Rees}, D.~E., \& {Semel}, M. 1991,
  \JournalTitle{A\&A}, 250, 463

\bibitem[{{Brown}(2014)}]{Brown2014}
{Brown}, T.~M. 2014,
  \href{http://dx.doi.org/10.1088/0004-637X/789/2/101}{\JournalTitle{ApJ}, 789,
  101}

\bibitem[{{Cranmer} \& {Saar}(2011)}]{Cranmer2011}
{Cranmer}, S.~R., \& {Saar}, S.~H. 2011,
  \href{http://dx.doi.org/10.1088/0004-637X/741/1/54}{\JournalTitle{ApJ}, 741,
  54}

\bibitem[{{DeRosa} {et~al.}(2012){DeRosa}, {Brun}, \& {Hoeksema}}]{DeRosa2012}
{DeRosa}, M.~L., {Brun}, A.~S., \& {Hoeksema}, J.~T. 2012,
  \href{http://dx.doi.org/10.1088/0004-637X/757/1/96}{\JournalTitle{ApJ}, 757,
  96}

\bibitem[{{do Nascimento} {et~al.}(2016){do Nascimento}, {Vidotto}, {Petit},
  {Folsom}, {Castro}, {Marsden}, {Morin}, {Porto de Mello}, {Meibom},
  {Jeffers}, {Guinan}, \& {Ribas}}]{Nasciemento2016}
{do Nascimento}, Jr., J.-D., {Vidotto}, A.~A., {Petit}, P., {et~al.} 2016,
  \href{http://dx.doi.org/10.3847/2041-8205/820/1/L15}{\JournalTitle{ApJl},
  820, L15}

\bibitem[{{Donati} \& {Brown}(1997)}]{Donati1997}
{Donati}, J.-F., \& {Brown}, S.~F. 1997, \JournalTitle{A\&A}, 326, 1135

\bibitem[{{Donati} {et~al.}(2003){Donati}, {Collier Cameron}, {Semel},
  {Hussain}, {Petit}, {Carter}, {Marsden}, {Mengel}, {L{\'o}pez Ariste},
  {Jeffers}, \& {Rees}}]{Donati2003}
{Donati}, J.-F., {Collier Cameron}, A., {Semel}, M., {et~al.} 2003,
  \href{http://dx.doi.org/10.1046/j.1365-2966.2003.07031.x}{\JournalTitle{MNRAS},
  345, 1145}

\bibitem[{{Donati} {et~al.}(2006){Donati}, {Howarth}, {Jardine}, {Petit},
  {Catala}, {Landstreet}, {Bouret}, {Alecian}, {Barnes}, {Forveille},
  {Paletou}, \& {Manset}}]{Donati2006}
{Donati}, J.-F., {Howarth}, I.~D., {Jardine}, M.~M., {et~al.} 2006,
  \href{http://dx.doi.org/10.1111/j.1365-2966.2006.10558.x}{\JournalTitle{MNRAS},
  370, 629}

\bibitem[{{Donati} {et~al.}(2008){Donati}, {Morin}, {Petit}, {Delfosse},
  {Forveille}, {Auri{\`e}re}, {Cabanac}, {Dintrans}, {Fares}, {Gastine},
  {Jardine}, {Ligni{\`e}res}, {Paletou}, {Ramirez Velez}, \&
  {Th{\'e}ado}}]{Donati2008}
{Donati}, J.-F., {Morin}, J., {Petit}, P., {et~al.} 2008,
  \href{http://dx.doi.org/10.1111/j.1365-2966.2008.13799.x}{\JournalTitle{MNRAS},
  390, 545}

\bibitem[{{Douglas} {et~al.}(2016){Douglas}, {Ag{\"u}eros}, {Covey}, {Cargile},
  {Barclay}, {Cody}, {Howell}, \& {Kopytova}}]{Douglas2016}
{Douglas}, S.~T., {Ag{\"u}eros}, M.~A., {Covey}, K.~R., {et~al.} 2016,
  \href{http://dx.doi.org/10.3847/0004-637X/822/1/47}{\JournalTitle{ApJ}, 822,
  47}

\bibitem[{{Douglas} {et~al.}(2017){Douglas}, {Ag{\"u}eros}, {Covey}, \&
  {Kraus}}]{Douglas2017}
{Douglas}, S.~T., {Ag{\"u}eros}, M.~A., {Covey}, K.~R., \& {Kraus}, A. 2017,
  \href{http://dx.doi.org/10.3847/1538-4357/aa6e52}{\JournalTitle{ApJ}, 842,
  83}

\bibitem[{{Fares} {et~al.}(2013){Fares}, {Moutou}, {Donati}, {Catala},
  {Shkolnik}, {Jardine}, {Cameron}, \& {Deleuil}}]{Fares2013}
{Fares}, R., {Moutou}, C., {Donati}, J.-F., {et~al.} 2013,
  \href{http://dx.doi.org/10.1093/mnras/stt1386}{\JournalTitle{MNRAS}, 435,
  1451}

\bibitem[{{Fares} {et~al.}(2009){Fares}, {Donati}, {Moutou}, {Bohlender},
  {Catala}, {Deleuil}, {Shkolnik}, {Collier Cameron}, {Jardine}, \&
  {Walker}}]{Fares2009}
{Fares}, R., {Donati}, J.-F., {Moutou}, C., {et~al.} 2009,
  \href{http://dx.doi.org/10.1111/j.1365-2966.2009.15303.x}{\JournalTitle{MNRAS},
  398, 1383}

\bibitem[{{Fares} {et~al.}(2010){Fares}, {Donati}, {Moutou}, {Jardine},
  {Grie{\ss}meier}, {Zarka}, {Shkolnik}, {Bohlender}, {Catala}, \& {Collier
  Cameron}}]{Fares2010}
---. 2010,
  \href{http://dx.doi.org/10.1111/j.1365-2966.2010.16715.x}{\JournalTitle{MNRAS},
  406, 409}

\bibitem[{{Fares} {et~al.}(2012){Fares}, {Donati}, {Moutou}, {Jardine},
  {Cameron}, {Lanza}, {Bohlender}, {Dieters}, {Mart{\'{\i}}nez Fiorenzano},
  {Maggio}, {Pagano}, \& {Shkolnik}}]{Fares2012}
---. 2012,
  \href{http://dx.doi.org/10.1111/j.1365-2966.2012.20780.x}{\JournalTitle{MNRAS},
  423, 1006}

\bibitem[{{Finley} \& {Matt}(2017)}]{Finley2017}
{Finley}, A.~J., \& {Matt}, S.~P. 2017,
  \href{http://dx.doi.org/10.3847/1538-4357/aa7fb9}{\JournalTitle{ApJ}, 845,
  46}

\bibitem[{{Finley} \& {Matt}(2018)}]{Finley2018}
---. 2018,
  \href{http://dx.doi.org/10.3847/1538-4357/aaaab5}{\JournalTitle{ApJ}, 854,
  78}

\bibitem[{{Finley} {et~al.}(2018){Finley}, {Matt}, \& {See}}]{Finley2018Sun}
{Finley}, A.~J., {Matt}, S.~P., \& {See}, V. 2018,
  \href{http://dx.doi.org/10.3847/1538-4357/aad7b6}{\JournalTitle{ApJ}, 864,
  125}

\bibitem[{{Finley} {et~al.}(2019){Finley}, {See}, \& {Matt}}]{Finley2019}
{Finley}, A.~J., {See}, V., \& {Matt}, S.~P. 2019,
  \href{http://dx.doi.org/10.3847/1538-4357/ab12d2}{\JournalTitle{ApJ}, 876,
  44}

\bibitem[{{Folsom} {et~al.}(2016){Folsom}, {Petit}, {Bouvier}, {L{\`e}bre},
  {Amard}, {Palacios}, {Morin}, {Donati}, {Jeffers}, {Marsden}, \&
  {Vidotto}}]{Folsom2016}
{Folsom}, C.~P., {Petit}, P., {Bouvier}, J., {et~al.} 2016,
  \href{http://dx.doi.org/10.1093/mnras/stv2924}{\JournalTitle{MNRAS}, 457,
  580}

\bibitem[{{Folsom} {et~al.}(2018{\natexlab{a}}){Folsom}, {Fossati}, {Wood},
  {Sreejith}, {Cubillos}, {Vidotto}, {Alecian}, {Girish}, {Lichtenegger},
  {Murthy}, {Petit}, \& {Valyavin}}]{Folsom2018Single}
{Folsom}, C.~P., {Fossati}, L., {Wood}, B.~E., {et~al.} 2018{\natexlab{a}},
  \JournalTitle{ArXiv e-prints},
  \href{http://arxiv.org/abs/1808.00406}{{\sffamily arXiv:1808.00406
  [astro-ph.SR]}}

\bibitem[{{Folsom} {et~al.}(2018{\natexlab{b}}){Folsom}, {Bouvier}, {Petit},
  {L{\`e}bre}, {Amard}, {Palacios}, {Morin}, {Donati}, \&
  {Vidotto}}]{Folsom2018}
{Folsom}, C.~P., {Bouvier}, J., {Petit}, P., {et~al.} 2018{\natexlab{b}},
  \href{http://dx.doi.org/10.1093/mnras/stx3021}{\JournalTitle{MNRAS}, 474,
  4956}

\bibitem[{{Gallet} \& {Bouvier}(2013)}]{Gallet2013}
{Gallet}, F., \& {Bouvier}, J. 2013,
  \href{http://dx.doi.org/10.1051/0004-6361/201321302}{\JournalTitle{A\&A},
  556, A36}

\bibitem[{{Gallet} \& {Bouvier}(2015)}]{Gallet2015}
---. 2015,
  \href{http://dx.doi.org/10.1051/0004-6361/201525660}{\JournalTitle{A\&A},
  577, A98}

\bibitem[{{Garraffo} {et~al.}(2015){Garraffo}, {Drake}, \&
  {Cohen}}]{Garraffo2015}
{Garraffo}, C., {Drake}, J.~J., \& {Cohen}, O. 2015,
  \href{http://dx.doi.org/10.1088/2041-8205/807/1/L6}{\JournalTitle{ApJl}, 807,
  L6}

\bibitem[{{Garraffo} {et~al.}(2016){Garraffo}, {Drake}, \&
  {Cohen}}]{Garraffo2016}
---. 2016,
  \href{http://dx.doi.org/10.1051/0004-6361/201628367}{\JournalTitle{A\&A},
  595, A110}

\bibitem[{{Garraffo} {et~al.}(2018){Garraffo}, {Drake}, {Dotter}, {Choi},
  {Burke}, {Moschou}, {Alvarado-Gomez}, {Kashyap}, \& {Cohen}}]{Garraffo2018}
{Garraffo}, C., {Drake}, J.~J., {Dotter}, A., {et~al.} 2018,
  \JournalTitle{ArXiv e-prints},
  \href{http://arxiv.org/abs/1804.01986}{{\sffamily arXiv:1804.01986
  [astro-ph.SR]}}

\bibitem[{{Gondoin}(2017)}]{Gondoin2017}
{Gondoin}, P. 2017,
  \href{http://dx.doi.org/10.1051/0004-6361/201629760}{\JournalTitle{A\&A},
  599, A122}

\bibitem[{{H{\'e}brard} {et~al.}(2016){H{\'e}brard}, {Donati}, {Delfosse},
  {Morin}, {Moutou}, \& {Boisse}}]{Hebrard2016}
{H{\'e}brard}, {\'E}.~M., {Donati}, J.-F., {Delfosse}, X., {et~al.} 2016,
  \href{http://dx.doi.org/10.1093/mnras/stw1346}{\JournalTitle{MNRAS}, 461,
  1465}

\bibitem[{{Jardine} \& {Collier Cameron}(2018)}]{Jardine2018}
{Jardine}, M., \& {Collier Cameron}, A. 2018,
  \href{http://dx.doi.org/10.1093/mnras/sty2872}{\JournalTitle{MNRAS}},
  \href{http://arxiv.org/abs/1810.09319}{{\sffamily arXiv:1810.09319
  [astro-ph.SR]}}

\bibitem[{{Jardine} {et~al.}(2017){Jardine}, {Vidotto}, \& {See}}]{Jardine2017}
{Jardine}, M., {Vidotto}, A.~A., \& {See}, V. 2017,
  \href{http://dx.doi.org/10.1093/mnrasl/slw206}{\JournalTitle{MNRAS}, 465,
  L25}

\bibitem[{{Jeffers} {et~al.}(2014){Jeffers}, {Petit}, {Marsden}, {Morin},
  {Donati}, \& {Folsom}}]{Jeffers2014}
{Jeffers}, S.~V., {Petit}, P., {Marsden}, S.~C., {et~al.} 2014,
  \href{http://dx.doi.org/10.1051/0004-6361/201423725}{\JournalTitle{A\&A},
  569, A79}

\bibitem[{{Johnstone} {et~al.}(2015){Johnstone}, {G{\"u}del}, {Brott}, \&
  {L{\"u}ftinger}}]{Johnstone2015}
{Johnstone}, C.~P., {G{\"u}del}, M., {Brott}, I., \& {L{\"u}ftinger}, T. 2015,
  \href{http://dx.doi.org/10.1051/0004-6361/201425301}{\JournalTitle{A\&A},
  577, A28}

\bibitem[{{Lavail} {et~al.}(2018){Lavail}, {Kochukhov}, \& {Wade}}]{Lavail2018}
{Lavail}, A., {Kochukhov}, O., \& {Wade}, G.~A. 2018,
  \href{http://dx.doi.org/10.1093/mnras/sty1825}{\JournalTitle{MNRAS}, 479,
  4836}

\bibitem[{{Lehmann} {et~al.}(2018){Lehmann}, {Hussain}, {Jardine}, {Mackay}, \&
  {Vidotto}}]{Lehmann2018ZDI}
{Lehmann}, L.~T., {Hussain}, G.~A.~J., {Jardine}, M.~M., {Mackay}, D.~H., \&
  {Vidotto}, A.~A. 2018, \JournalTitle{ArXiv e-prints},
  \href{http://arxiv.org/abs/1811.03703}{{\sffamily arXiv:1811.03703
  [astro-ph.SR]}}

\bibitem[{{Linker} {et~al.}(2017){Linker}, {Caplan}, {Downs}, {Riley}, {Mikic},
  {Lionello}, {Henney}, {Arge}, {Liu}, {Derosa}, {Yeates}, \&
  {Owens}}]{Linker2017}
{Linker}, J.~A., {Caplan}, R.~M., {Downs}, C., {et~al.} 2017,
  \href{http://dx.doi.org/10.3847/1538-4357/aa8a70}{\JournalTitle{ApJ}, 848,
  70}

\bibitem[{{Marsden} {et~al.}(2011){Marsden}, {Jardine}, {Ram{\'{\i}}rez
  V{\'e}lez}, {Alecian}, {Brown}, {Carter}, {Donati}, {Dunstone}, {Hart},
  {Semel}, \& {Waite}}]{Marsden2011}
{Marsden}, S.~C., {Jardine}, M.~M., {Ram{\'{\i}}rez V{\'e}lez}, J.~C., {et~al.}
  2011,
  \href{http://dx.doi.org/10.1111/j.1365-2966.2011.18367.x}{\JournalTitle{MNRAS},
  413, 1922}

\bibitem[{{Matt} {et~al.}(2015){Matt}, {Brun}, {Baraffe}, {Bouvier}, \&
  {Chabrier}}]{Matt2015}
{Matt}, S.~P., {Brun}, A.~S., {Baraffe}, I., {Bouvier}, J., \& {Chabrier}, G.
  2015,
  \href{http://dx.doi.org/10.1088/2041-8205/799/2/L23}{\JournalTitle{ApJl},
  799, L23}

\bibitem[{{Mengel} {et~al.}(2016){Mengel}, {Fares}, {Marsden}, {Carter},
  {Jeffers}, {Petit}, {Donati}, {Folsom}, \& {BCool
  Collaboration}}]{Mengel2016}
{Mengel}, M.~W., {Fares}, R., {Marsden}, S.~C., {et~al.} 2016,
  \href{http://dx.doi.org/10.1093/mnras/stw828}{\JournalTitle{MNRAS}, 459,
  4325}

\bibitem[{{Morgenthaler} {et~al.}(2012){Morgenthaler}, {Petit}, {Saar},
  {Solanki}, {Morin}, {Marsden}, {Auri{\`e}re}, {Dintrans}, {Fares}, {Gastine},
  {Lanoux}, {Ligni{\`e}res}, {Paletou}, {Ram{\'{\i}}rez V{\'e}lez},
  {Th{\'e}ado}, \& {Van Grootel}}]{Morgenthaler2012}
{Morgenthaler}, A., {Petit}, P., {Saar}, S., {et~al.} 2012,
  \href{http://dx.doi.org/10.1051/0004-6361/201118139}{\JournalTitle{A\& A},
  540, A138}

\bibitem[{{Morin} {et~al.}(2010){Morin}, {Donati}, {Petit}, {Delfosse},
  {Forveille}, \& {Jardine}}]{Morin2010}
{Morin}, J., {Donati}, J.-F., {Petit}, P., {et~al.} 2010,
  \href{http://dx.doi.org/10.1111/j.1365-2966.2010.17101.x}{\JournalTitle{MNRAS},
  407, 2269}

\bibitem[{{Morin} {et~al.}(2008{\natexlab{a}}){Morin}, {Donati}, {Petit},
  {Delfosse}, {Forveille}, {Albert}, {Auri{\`e}re}, {Cabanac}, {Dintrans},
  {Fares}, {Gastine}, {Jardine}, {Ligni{\`e}res}, {Paletou}, {Ramirez Velez},
  \& {Th{\'e}ado}}]{Morin2008}
---. 2008{\natexlab{a}},
  \href{http://dx.doi.org/10.1111/j.1365-2966.2008.13809.x}{\JournalTitle{MNRAS},
  390, 567}

\bibitem[{{Morin} {et~al.}(2008{\natexlab{b}}){Morin}, {Donati}, {Forveille},
  {Delfosse}, {Dobler}, {Petit}, {Jardine}, {Collier Cameron}, {Albert},
  {Manset}, {Dintrans}, {Chabrier}, \& {Valenti}}]{Morin2008V374Peg}
{Morin}, J., {Donati}, J.-F., {Forveille}, T., {et~al.} 2008{\natexlab{b}},
  \href{http://dx.doi.org/10.1111/j.1365-2966.2007.12709.x}{\JournalTitle{MNRAS},
  384, 77}

\bibitem[{{Pantolmos} \& {Matt}(2017)}]{Pantolmos2017}
{Pantolmos}, G., \& {Matt}, S.~P. 2017,
  \href{http://dx.doi.org/10.3847/1538-4357/aa9061}{\JournalTitle{ApJ}, 849,
  83}

\bibitem[{{Perri} {et~al.}(2018){Perri}, {Brun}, {R{\'e}ville}, \&
  {Strugarek}}]{Perri2018}
{Perri}, B., {Brun}, A.~S., {R{\'e}ville}, V., \& {Strugarek}, A. 2018,
  \href{http://dx.doi.org/10.1017/S0022377818000880}{\JournalTitle{Journal of
  Plasma Physics}, 84, 765840501}

\bibitem[{{Petit} {et~al.}(2008){Petit}, {Dintrans}, {Solanki}, {Donati},
  {Auri{\`e}re}, {Ligni{\`e}res}, {Morin}, {Paletou}, {Ramirez Velez},
  {Catala}, \& {Fares}}]{Petit2008}
{Petit}, P., {Dintrans}, B., {Solanki}, S.~K., {et~al.} 2008,
  \href{http://dx.doi.org/10.1111/j.1365-2966.2008.13411.x}{\JournalTitle{MNRAS},
  388, 80}

\bibitem[{{Pinto} {et~al.}(2011){Pinto}, {Brun}, {Jouve}, \&
  {Grappin}}]{Pinto2011}
{Pinto}, R.~F., {Brun}, A.~S., {Jouve}, L., \& {Grappin}, R. 2011,
  \href{http://dx.doi.org/10.1088/0004-637X/737/2/72}{\JournalTitle{ApJ}, 737,
  72}

\bibitem[{{Pizzolato} {et~al.}(2003){Pizzolato}, {Maggio}, {Micela},
  {Sciortino}, \& {Ventura}}]{Pizzolato2003}
{Pizzolato}, N., {Maggio}, A., {Micela}, G., {Sciortino}, S., \& {Ventura}, P.
  2003, \href{http://dx.doi.org/10.1051/0004-6361:20021560}{\JournalTitle{aap},
  397, 147}

\bibitem[{{Reiners} \& {Basri}(2009)}]{Reiners2009}
{Reiners}, A., \& {Basri}, G. 2009,
  \href{http://dx.doi.org/10.1051/0004-6361:200811450}{\JournalTitle{A\&A},
  496, 787}

\bibitem[{{R{\'e}ville} \& {Brun}(2017)}]{Reville2017}
{R{\'e}ville}, V., \& {Brun}, A.~S. 2017,
  \href{http://dx.doi.org/10.3847/1538-4357/aa9218}{\JournalTitle{ApJ}, 850,
  45}

\bibitem[{{R{\'e}ville} {et~al.}(2015){R{\'e}ville}, {Brun}, {Matt},
  {Strugarek}, \& {Pinto}}]{Reville2015}
{R{\'e}ville}, V., {Brun}, A.~S., {Matt}, S.~P., {Strugarek}, A., \& {Pinto},
  R.~F. 2015,
  \href{http://dx.doi.org/10.1088/0004-637X/798/2/116}{\JournalTitle{ApJ}, 798,
  116}

\bibitem[{{Sadeghi Ardestani} {et~al.}(2017){Sadeghi Ardestani}, {Guillot}, \&
  {Morel}}]{Ardestani2017}
{Sadeghi Ardestani}, L., {Guillot}, T., \& {Morel}, P. 2017,
  \href{http://dx.doi.org/10.1093/mnras/stx2039}{\JournalTitle{MNRAS}, 472,
  2590}

\bibitem[{{See} {et~al.}(2017){See}, {Jardine}, {Vidotto}, {Donati}, {Boro
  Saikia}, {Fares}, {Folsom}, {H{\'e}brard}, {Jeffers}, {Marsden}, {Morin},
  {Petit}, {Waite}, \& {BCool Collaboration}}]{See2017}
{See}, V., {Jardine}, M., {Vidotto}, A.~A., {et~al.} 2017,
  \href{http://dx.doi.org/10.1093/mnras/stw3094}{\JournalTitle{MNRAS}, 466,
  1542}

\bibitem[{{See} {et~al.}(2018){See}, {Jardine}, {Vidotto}, {Donati}, {Boro
  Saikia}, {Fares}, {Folsom}, {Jeffers}, {Marsden}, {Morin}, {Petit}, \& {BCool
  Collaboration}}]{See2018}
---. 2018, \href{http://dx.doi.org/10.1093/mnras/stx2599}{\JournalTitle{MNRAS},
  474, 536}

\bibitem[{{See} {et~al.}(2019){See}, {Matt}, {Folsom}, {Boro Saikia}, {Donati},
  {Fares}, {Finley}, {H{\'e}brard}, {Jardine}, {Jeffers}, {Lehmann}, {Marsden},
  {Mengel}, {Morin}, {Petit}, {Vidotto}, {Waite}, \& {The BCool
  Collaboration}}]{See2019}
{See}, V., {Matt}, S.~P., {Folsom}, C.~P., {et~al.} 2019,
  \href{http://dx.doi.org/10.3847/1538-4357/ab1096}{\JournalTitle{ApJ}, 876,
  118}

\bibitem[{{Semel}(1989)}]{Semel1989}
{Semel}, M. 1989, \JournalTitle{A\&A}, 225, 456

\bibitem[{{Skumanich}(1972)}]{Skumanich1972}
{Skumanich}, A. 1972,
  \href{http://dx.doi.org/10.1086/151310}{\JournalTitle{ApJ}, 171, 565}

\bibitem[{{Stelzer} {et~al.}(2016){Stelzer}, {Damasso}, {Scholz}, \&
  {Matt}}]{Stelzer2016}
{Stelzer}, B., {Damasso}, M., {Scholz}, A., \& {Matt}, S.~P. 2016,
  \href{http://dx.doi.org/10.1093/mnras/stw1936}{\JournalTitle{MNRAS}, 463,
  1844}

\bibitem[{{Suzuki} {et~al.}(2013){Suzuki}, {Imada}, {Kataoka}, {Kato},
  {Matsumoto}, {Miyahara}, \& {Tsuneta}}]{Suzuki2013}
{Suzuki}, T.~K., {Imada}, S., {Kataoka}, R., {et~al.} 2013,
  \href{http://dx.doi.org/10.1093/pasj/65.5.98}{\JournalTitle{PASJ}, 65, 98}

\bibitem[{{van Saders} {et~al.}(2016){van Saders}, {Ceillier}, {Metcalfe},
  {Silva Aguirre}, {Pinsonneault}, {Garc{\'{\i}}a}, {Mathur}, \&
  {Davies}}]{vanSaders2016}
{van Saders}, J.~L., {Ceillier}, T., {Metcalfe}, T.~S., {et~al.} 2016,
  \href{http://dx.doi.org/10.1038/nature16168}{\JournalTitle{Nature}, 529, 181}

\bibitem[{{Vidotto} \& {Bourrier}(2017)}]{Vidotto2017}
{Vidotto}, A.~A., \& {Bourrier}, V. 2017,
  \href{http://dx.doi.org/10.1093/mnras/stx1543}{\JournalTitle{MNRAS}, 470,
  4026}

\bibitem[{{Vidotto} {et~al.}(2014){Vidotto}, {Jardine}, {Morin}, {Donati},
  {Opher}, \& {Gombosi}}]{Vidotto2014Torque}
{Vidotto}, A.~A., {Jardine}, M., {Morin}, J., {et~al.} 2014,
  \href{http://dx.doi.org/10.1093/mnras/stt2265}{\JournalTitle{MNRAS}, 438,
  1162}

\bibitem[{{Vidotto} {et~al.}(2018){Vidotto}, {Lehmann}, {Jardine}, \&
  {Pevtsov}}]{Vidotto2018}
{Vidotto}, A.~A., {Lehmann}, L.~T., {Jardine}, M., \& {Pevtsov}, A.~A. 2018,
  \href{http://dx.doi.org/10.1093/mnras/sty1926}{\JournalTitle{MNRAS}, 480,
  477}

\bibitem[{{Vidotto} {et~al.}(2016){Vidotto}, {Donati}, {Jardine}, {See},
  {Petit}, {Boisse}, {Boro Saikia}, {H{\'e}brard}, {Jeffers}, {Marsden}, \&
  {Morin}}]{Vidotto2016}
{Vidotto}, A.~A., {Donati}, J.-F., {Jardine}, M., {et~al.} 2016,
  \href{http://dx.doi.org/10.1093/mnrasl/slv147}{\JournalTitle{MNRAS}, 455,
  L52}

\bibitem[{{Waite} {et~al.}(2015){Waite}, {Marsden}, {Carter}, {Petit},
  {Donati}, {Jeffers}, \& {Boro Saikia}}]{Waite2015}
{Waite}, I.~A., {Marsden}, S.~C., {Carter}, B.~D., {et~al.} 2015,
  \href{http://dx.doi.org/10.1093/mnras/stv006}{\JournalTitle{MNRAS}, 449, 8}

\bibitem[{{Waite} {et~al.}(2017){Waite}, {Marsden}, {Carter}, {Petit},
  {Jeffers}, {Morin}, {Vidotto}, {Donati}, \& {BCool
  Collaboration}}]{Waite2017}
---. 2017, \href{http://dx.doi.org/10.1093/mnras/stw2731}{\JournalTitle{MNRAS},
  465, 2076}

\bibitem[{{Weber} \& {Davis}(1967)}]{Weber1967}
{Weber}, E.~J., \& {Davis}, Jr., L. 1967,
  \href{http://dx.doi.org/10.1086/149138}{\JournalTitle{ApJ}, 148, 217}

\bibitem[{{Wood} {et~al.}(2014){Wood}, {M{\"u}ller}, {Redfield}, \&
  {Edelman}}]{Wood2014}
{Wood}, B.~E., {M{\"u}ller}, H.-R., {Redfield}, S., \& {Edelman}, E. 2014,
  \href{http://dx.doi.org/10.1088/2041-8205/781/2/L33}{\JournalTitle{ApJ}, 781,
  L33}

\bibitem[{{Wright} \& {Drake}(2016)}]{Wright2016}
{Wright}, N.~J., \& {Drake}, J.~J. 2016,
  \href{http://dx.doi.org/10.1038/nature18638}{\JournalTitle{Nature}, 535, 526}

\bibitem[{{Wright} {et~al.}(2011){Wright}, {Drake}, {Mamajek}, \&
  {Henry}}]{Wright2011}
{Wright}, N.~J., {Drake}, J.~J., {Mamajek}, E.~E., \& {Henry}, G.~W. 2011,
  \href{http://dx.doi.org/10.1088/0004-637X/743/1/48}{\JournalTitle{ApJ}, 743,
  48}

\bibitem[{{Wright} {et~al.}(2018){Wright}, {Newton}, {Williams}, {Drake}, \&
  {Yadav}}]{Wright2018}
{Wright}, N.~J., {Newton}, E.~R., {Williams}, P.~K.~G., {Drake}, J.~J., \&
  {Yadav}, R.~K. 2018,
  \href{http://dx.doi.org/10.1093/mnras/sty1670}{\JournalTitle{MNRAS}, 479,
  2351}

\end{thebibliography}

\appendix

\begin{sidewaystable}
\caption{Numerical values used and derived in this study for our sample of stars. Listed are the dipole, quadrupole \& octupole field strengths, mass-loss rates, spin-down torque, spin-down time-scale and the paper in which the ZDI map of each star was originally published. For quantities with multiple estimates, the method used is noted in brackets.}
\label{tab:Values}
\center
\begin{tabular}{lcccccccccccc}
\hline\hline
Star ID & $\langle B_{\rm d} \rangle$ & $\langle B_{\rm q} \rangle$ & $\langle B_{\rm o} \rangle$ & $\dot{M}$ (CS11) & $\dot{M}$ (M15) & $\dot{M}$ (mod M15) & $\dot{M}_{\rm crit}$ & $T$ (CS11) & $T$ (M15) & $\tau$ (CS11) & $\tau$ (M15) & Reference\\
 & (G) & (G) & (G) & ($M_{\odot}\ {\rm yr}^{-1}$) & ($M_{\odot}\ {\rm yr}^{-1}$) & ($M_{\odot}\ {\rm yr}^{-1}$) & ($M_{\odot}\ {\rm yr}^{-1}$) & (erg) & (erg) & (Gyr) & (Gyr) & \\
\hline
\textbf{Solar-like stars} \\
HD 3651	&	2.9	&	1.93	&	0.701	&	4.72E-15	&	1.08E-12	&	3.32E-15	&	6.34E-13	&	1.32E+29	&	2.78E+30	&	224	&	10.7	&	Petit et al. (in prep)\\
HD 9986	&	0.539	&	0.291	&	0.114	&	5.63E-14	&	4.87E-12	&	9.03E-14	&	3.85E-14	&	3.60E+29	&	1.29E+31	&	201	&	5.6	&	Petit et al. (in prep)	\\
HD 10476	&	1.62	&	1.39	&	0.901	&	4.02E-14	&	1.59E-11	&	2.79E-13	&	1.26E-13	&	5.47E+29	&	4.61E+31	&	130	&	1.54	&	Petit et al. (in prep)	\\
$\kappa$ Cet	&	11.4	&	7.61	&	4.44	&	3.36E-13	&	1.28E-11	&	3.49E-14	&	1.08E-11	&	2.67E+31	&	1.99E+32	&	6.81	&	0.915	&	\citet{Nasciemento2016}	\\
$\epsilon$ Eri (2007)	&	11	&	2.93	&	1.47	&	5.05E-14	&	2.86E-11	&	9.93E-14	&	1.28E-11	&	3.97E+30	&	1.45E+32	&	30.3	&	0.827	&	\citet{Jeffers2014}	\\
$\epsilon$ Eri (2008)	&	7.48	&	4.85	&	2.82	&	5.05E-14	&	2.87E-11	&	1.90E-13	&	2.84E-12	&	2.80E+30	&	1.45E+32	&	43	&	0.827	&	\citet{Jeffers2014}	\\
$\epsilon$ Eri (2010)	&	6.47	&	6.36	&	4.65	&	5.05E-14	&	2.71E-11	&	2.42E-13	&	1.26E-12	&	2.45E+30	&	1.45E+32	&	49.1	&	0.827	&	\citet{Jeffers2014}	\\
$\epsilon$ Eri (2011)	&	6.79	&	3.24	&	2.68	&	5.05E-14	&	3.09E-11	&	2.23E-13	&	3.20E-12	&	2.56E+30	&	1.45E+32	&	46.9	&	0.827	&	\citet{Jeffers2014}	\\
$\epsilon$ Eri (2012)	&	12.5	&	5.5	&	3.85	&	5.05E-14	&	2.40E-11	&	7.94E-14	&	1.17E-11	&	4.49E+30	&	1.45E+32	&	26.8	&	0.827	&	\citet{Jeffers2014}	\\
$\epsilon$ Eri (2013)	&	15.6	&	4.03	&	4.06	&	5.05E-14	&	2.14E-11	&	5.48E-14	&	2.64E-11	&	5.48E+30	&	1.45E+32	&	21.9	&	0.827	&	\citet{Jeffers2014}	\\
HD 39587	&	5.37	&	5.76	&	6.05	&	1.93E-12	&	6.32E-11	&	7.93E-13	&	1.69E-12	&	9.41E+31	&	1.43E+33	&	3.72	&	0.244	&	Petit et al. (in prep)	\\
HD 56124	&	1.94	&	0.708	&	0.235	&	9.12E-14	&	4.50E-12	&	2.63E-14	&	6.52E-13	&	1.64E+30	&	2.12E+31	&	57.4	&	4.44	&	Petit et al. (in prep)	\\
HD 72905	&	6.69	&	5.57	&	5.74	&	1.60E-12	&	4.06E-11	&	3.03E-13	&	3.33E-12	&	8.45E+31	&	8.69E+32	&	3.8	&	0.369	&	Petit et al. (in prep)	\\
HD 73350	&	4.94	&	4.3	&	4.63	&	2.13E-13	&	7.36E-12	&	3.29E-14	&	1.60E-12	&	8.11E+30	&	7.48E+31	&	17.3	&	1.87	&	\citet{Petit2008}	\\
HD 75332	&	5.14	&	1.91	&	1.86	&	1.31E-12	&	1.01E-11	&	2.96E-14	&	6.91E-12	&	1.16E+32	&	3.79E+32	&	4	&	1.23	&	Petit et al. (in prep)	\\
HD 76151	&	2.71	&	1.32	&	0.608	&	6.42E-14	&	3.74E-12	&	1.57E-14	&	9.63E-13	&	1.56E+30	&	1.84E+31	&	55.3	&	4.68	&	\citet{Petit2008}	\\
HD 78366	&	10.4	&	8.25	&	4.03	&	2.97E-13	&	2.14E-12	&	5.49E-15	&	9.40E-12	&	2.61E+31	&	7.61E+31	&	6.69	&	2.29	&	Petit et al. (in prep)	\\
HD 101501	&	7.61	&	6.48	&	2.57	&	6.09E-14	&	4.25E-12	&	1.17E-14	&	3.48E-12	&	3.42E+30	&	3.55E+31	&	20.2	&	1.94	&	Petit et al. (in prep)	\\
$\xi$ Boo A (2007)	&	21.9	&	10.1	&	7.99	&	2.76E-13	&	3.95E-11	&	1.01E-13	&	4.53E-11	&	4.42E+31	&	6.52E+32	&	5.02	&	0.341	&	\citet{Morgenthaler2012}	\\
$\xi$ Boo A (2008)	&	12.1	&	7.86	&	7.88	&	2.76E-13	&	5.33E-11	&	2.75E-13	&	9.87E-12	&	2.58E+31	&	6.52E+32	&	8.63	&	0.341	&	\citet{Morgenthaler2012}	\\
$\xi$ Boo A (2009)	&	14.6	&	10.9	&	10.4	&	2.76E-13	&	4.79E-11	&	2.00E-13	&	1.23E-11	&	3.06E+31	&	6.52E+32	&	7.26	&	0.341	&	\citet{Morgenthaler2012}	\\
$\xi$ Boo A (Jan 2010)	&	9.04	&	5.43	&	4.73	&	2.76E-13	&	6.23E-11	&	4.51E-13	&	5.96E-12	&	1.97E+31	&	6.52E+32	&	11.3	&	0.341	&	\citet{Morgenthaler2012}	\\
$\xi$ Boo A (Jun 2010)	&	15.2	&	8.01	&	5.64	&	2.76E-13	&	5.17E-11	&	1.87E-13	&	1.93E-11	&	3.17E+31	&	6.52E+32	&	7.01	&	0.341	&	\citet{Morgenthaler2012}	\\
$\xi$ Boo A (Jul 2010)	&	11.2	&	6.34	&	5.04	&	2.76E-13	&	5.83E-11	&	3.16E-13	&	9.60E-12	&	2.39E+31	&	6.52E+32	&	9.32	&	0.341	&	\citet{Morgenthaler2012}	\\
$\xi$ Boo A (2011)	&	14.4	&	6.97	&	3.59	&	2.76E-13	&	5.48E-11	&	2.04E-13	&	1.89E-11	&	3.02E+31	&	6.52E+32	&	7.35	&	0.341	&	\citet{Morgenthaler2012}	\\
$\xi$ Boo B	&	9.34	&	8.15	&	6.34	&	2.70E-15	&	4.20E-11	&	4.35E-13	&	1.63E-12	&	2.89E+29	&	1.15E+32	&	284	&	0.714	&	Petit et al. (in prep)	\\
18 Sco	&	0.776	&	0.92	&	0.407	&	6.17E-14	&	3.98E-12	&	6.22E-14	&	2.99E-14	&	5.73E+29	&	1.46E+31	&	126	&	4.92	&	\citet{Petit2008}	\\
HD 166435	&	8.64	&	8.75	&	6.04	&	2.25E-12	&	1.17E-10	&	1.20E-12	&	4.06E-12	&	1.80E+32	&	3.24E+33	&	2.79	&	0.155	&	Petit et al. (in prep)	\\
HD 175726	&	4.21	&	4.1	&	3.14	&	2.99E-12	&	7.72E-11	&	1.23E-12	&	1.20E-12	&	1.39E+32	&	1.85E+33	&	3.25	&	0.245	&	Petit et al. (in prep)	\\
HD 190771	&	6.24	&	3.61	&	3.92	&	4.96E-13	&	1.84E-11	&	8.62E-14	&	4.38E-12	&	2.42E+31	&	2.38E+32	&	8.39	&	0.854	&	\citet{Petit2008}	\\
61 Cyg A (2007)	&	9.77	&	4.76	&	2.62	&	8.03E-16	&	1.18E-12	&	3.03E-15	&	4.75E-12	&	6.94E+28	&	3.62E+30	&	319	&	6.11	&	\citet{Saikia2016}	\\
61 Cyg A (2008)	&	2.16	&	1.8	&	0.85	&	8.03E-16	&	3.73E-12	&	3.88E-14	&	1.29E-13	&	1.74E+28	&	3.62E+30	&	1.27E+03	&	6.11	&	\citet{Saikia2016}	\\
61 Cyg A (2010)	&	2.38	&	2.59	&	3.59	&	8.03E-16	&	2.93E-12	&	3.31E-14	&	1.08E-13	&	1.90E+28	&	3.62E+30	&	1.16E+03	&	6.11	&	\citet{Saikia2016}	\\
61 Cyg A (2013)	&	8.33	&	3.79	&	1.73	&	8.03E-16	&	1.55E-12	&	3.97E-15	&	3.68E-12	&	6.00E+28	&	3.62E+30	&	369	&	6.11	&	\citet{Saikia2016}	\\
61 Cyg A (2014)	&	6.91	&	3.33	&	1.56	&	8.03E-16	&	2.12E-12	&	5.44E-15	&	2.40E-12	&	5.05E+28	&	3.62E+30	&	438	&	6.11	&	\citet{Saikia2016}	\\
61 Cyg A (Aug 2015)	&	10.4	&	4.05	&	1.83	&	8.03E-16	&	1.06E-12	&	2.71E-15	&	6.58E-12	&	7.37E+28	&	3.62E+30	&	300	&	6.11	&	\citet{Saikia2016}	\\
61 Cyg A (Oct 2015)	&	6.88	&	2.71	&	1.82	&	8.03E-16	&	2.14E-12	&	5.48E-15	&	2.83E-12	&	5.03E+28	&	3.62E+30	&	439	&	6.11	&	\citet{Saikia2018}	\\
61 Cyg A (Dec 2015)	&	5.84	&	2.66	&	1.32	&	8.03E-16	&	2.51E-12	&	7.22E-15	&	1.81E-12	&	4.34E+28	&	3.62E+30	&	510	&	6.11	&	\citet{Saikia2018}	\\
61 Cyg A (2016)	&	6.23	&	3.67	&	3.44	&	8.03E-16	&	2.24E-12	&	6.47E-15	&	1.61E-12	&	4.60E+28	&	3.62E+30	&	481	&	6.11	&	\citet{Saikia2018}	\\
61 Cyg A (Jul 2017)	&	6.62	&	1.93	&	0.672	&	8.03E-16	&	2.28E-12	&	5.84E-15	&	3.25E-12	&	4.86E+28	&	3.62E+30	&	455	&	6.11	&	\citet{Saikia2018}	\\
61 Cyg A (Dec 2017)	&	3.84	&	1.71	&	0.928	&	8.03E-16	&	3.29E-12	&	1.47E-14	&	7.97E-13	&	2.95E+28	&	3.62E+30	&	749	&	6.11	&	\citet{Saikia2018}	\\
61 Cyg A (2018)	&	8.74	&	3.61	&	2.32	&	8.03E-16	&	1.43E-12	&	3.66E-15	&	4.40E-12	&	6.27E+28	&	3.62E+30	&	353	&	6.11	&	\citet{Saikia2018}	\\
HN Peg (2007)	&	9.62	&	6.61	&	4.01	&	2.01E-12	&	4.98E-11	&	2.36E-13	&	9.12E-12	&	1.62E+32	&	1.29E+33	&	2.55	&	0.321	&	\citet{Saikia2015}	\\
HN Peg (2008)	&	6.27	&	4.09	&	4.35	&	2.01E-12	&	5.70E-11	&	4.86E-13	&	4.11E-12	&	1.09E+32	&	1.29E+33	&	3.78	&	0.321	&	\citet{Saikia2015}	\\
HN Peg (2009)	&	6.83	&	3.86	&	4.2	&	2.01E-12	&	5.67E-11	&	4.22E-13	&	5.67E-12	&	1.18E+32	&	1.29E+33	&	3.49	&	0.321	&	\citet{Saikia2015}	\\
HN Peg (2010)	&	9.15	&	6.17	&	5.82	&	2.01E-12	&	4.89E-11	&	2.57E-13	&	8.43E-12	&	1.55E+32	&	1.29E+33	&	2.67	&	0.321	&	\citet{Saikia2015}	\\
HN Peg (2011)	&	7.72	&	6.8	&	5.94	&	2.01E-12	&	4.96E-11	&	3.43E-13	&	4.32E-12	&	1.32E+32	&	1.29E+33	&	3.12	&	0.321	&	\citet{Saikia2015}	\\
HN Peg (2013)	&	13.7	&	8.04	&	4.56	&	2.01E-12	&	4.08E-11	&	1.31E-13	&	2.18E-11	&	2.23E+32	&	1.29E+33	&	1.85	&	0.321	&	\citet{Saikia2015}	\\
\hline
\end{tabular}
\end{sidewaystable}

\begin{sidewaystable}
\contcaption{continued}
\center
\begin{tabular}{lcccccccccccc}
\hline\hline
Star ID & $\langle B_{\rm d} \rangle$ & $\langle B_{\rm q} \rangle$ & $\langle B_{\rm o} \rangle$ & $\dot{M}$ (CS11) & $\dot{M}$ (M15) & $\dot{M}$ (mod M15) & $\dot{M}_{\rm crit}$ & $T$ (CS11) & $T$ (M15) & $\tau$ (CS11) & $\tau$ (M15) & Reference\\
 & (G) & (G) & (G) & ($M_{\odot}\ {\rm yr}^{-1}$) & ($M_{\odot}\ {\rm yr}^{-1}$) & ($M_{\odot}\ {\rm yr}^{-1}$) & ($M_{\odot}\ {\rm yr}^{-1}$) & (erg) & (erg) & (Gyr) & (Gyr) & \\
\hline
HD 219134	&	1.33	&	1.93	&	0.479	&	1.46E-15	&	2.83E-12	&	3.73E-14	&	3.46E-14	&	2.45E+28	&	3.65E+30	&	1070	&	7.2	&	\citet{Folsom2018Single}	\\
AV 1693	&	13.5	&	17.1	&	7.67	&	1.65E-13	&	1.49E-11	&	5.66E-14	&	4.89E-12	&	1.47E+31	&	2.09E+32	&	10.1	&	0.708	&	\citet{Folsom2018}	\\
AV 1826	&	9.5	&	10.2	&	7.37	&	8.85E-14	&	2.89E-11	&	1.67E-13	&	2.94E-12	&	6.67E+30	&	2.39E+32	&	19.5	&	0.544	&	\citet{Folsom2018}	\\
AV 2177	&	6.23	&	3.52	&	2.2	&	1.12E-13	&	3.40E-11	&	2.71E-13	&	2.54E-12	&	4.89E+30	&	2.00E+32	&	30.6	&	0.746	&	\citet{Folsom2018}	\\
AV 523	&	9.32	&	7.36	&	6.76	&	2.23E-14	&	2.83E-11	&	1.72E-13	&	3.39E-12	&	1.90E+30	&	1.46E+32	&	51.3	&	0.668	&	\citet{Folsom2018}	\\
EP Eri	&	14.9	&	9.55	&	1.61	&	7.68E-14	&	6.30E-11	&	3.01E-13	&	1.06E-11	&	9.23E+30	&	4.91E+32	&	19.4	&	0.365	&	\citet{Folsom2018}	\\
HH Leo	&	15.5	&	8.78	&	6.14	&	3.27E-13	&	5.74E-11	&	2.19E-13	&	1.84E-11	&	3.80E+31	&	7.77E+32	&	6.56	&	0.321	&	\citet{Folsom2018}	\\
Mel25-151	&	12	&	7.8	&	4.77	&	5.52E-14	&	2.99E-11	&	1.15E-13	&	9.37E-12	&	6.18E+30	&	2.34E+32	&	18.8	&	0.498	&	\citet{Folsom2018}	\\
Mel25-179	&	15.5	&	7.89	&	5.6	&	8.56E-14	&	2.77E-11	&	7.76E-14	&	2.15E-11	&	1.15E+31	&	2.76E+32	&	10.9	&	0.453	&	\citet{Folsom2018}	\\
Mel25-21	&	10.5	&	4.38	&	2.83	&	1.70E-13	&	2.88E-11	&	9.50E-14	&	1.41E-11	&	1.48E+31	&	2.74E+32	&	9.31	&	0.503	&	\citet{Folsom2018}	\\
Mel25-43	&	5.59	&	3.05	&	1.35	&	7.59E-14	&	3.79E-11	&	3.31E-13	&	2.26E-12	&	3.43E+30	&	1.93E+32	&	35.7	&	0.634	&	\citet{Folsom2018}	\\
Mel25-5	&	6.71	&	3.04	&	2.14	&	8.32E-14	&	4.45E-11	&	2.89E-13	&	5.51E-12	&	6.22E+30	&	3.10E+32	&	18.5	&	0.371	&	\citet{Folsom2018}	\\
TYC 1987-509-1	&	11.4	&	10.2	&	5.51	&	1.53E-13	&	1.72E-11	&	6.45E-14	&	5.83E-12	&	1.16E+31	&	1.85E+32	&	12.2	&	0.769	&	\citet{Folsom2018}	\\
V447 Lac	&	9.95	&	10.3	&	7.8	&	3.06E-13	&	1.38E-10	&	1.71E-12	&	3.39E-12	&	2.95E+31	&	1.90E+33	&	10.3	&	0.159	&	\citet{Folsom2016}	\\
DX Leo	&	26.1	&	11.7	&	6.01	&	3.11E-13	&	5.14E-11	&	1.32E-13	&	6.10E-11	&	5.93E+31	&	9.45E+32	&	4.21	&	0.264	&	\citet{Folsom2016}	\\
V439 And	&	9.51	&	4.61	&	2.46	&	4.35E-13	&	7.31E-11	&	4.29E-13	&	1.01E-11	&	3.59E+31	&	9.04E+32	&	6.59	&	0.262	&	\citet{Folsom2016}	\\
																									
\textbf{Young Suns} \\																									
AB Dor (2001)	&	68.1	&	103	&	94.8	&	7.81E-13	&	2.73E-11	&	6.99E-14	&	1.35E-10	&	4.23E+33	&	2.90E+34	&	0.621	&	0.0905	&	\citet{Donati2003}	\\
AB Dor (2002)	&	142	&	55.4	&	54.1	&	7.81E-13	&	7.87E-12	&	2.02E-14	&	3.11E-09	&	8.30E+33	&	2.90E+34	&	0.317	&	0.0905	&	\citet{Donati2003}	\\
BD-16351	&	34	&	23.6	&	13.8	&	4.42E-13	&	8.71E-11	&	2.28E-13	&	8.20E-11	&	1.99E+32	&	3.52E+33	&	2.1	&	0.119	&	\citet{Folsom2016}	\\
HII 296	&	65.4	&	32.5	&	27	&	3.23E-13	&	2.93E-11	&	7.52E-14	&	4.93E-10	&	4.47E+32	&	5.14E+33	&	1.15	&	0.0999	&	\citet{Folsom2016}	\\
HII 739	&	7.44	&	5.89	&	6.99	&	7.37E-12	&	2.46E-10	&	4.07E-12	&	4.84E-12	&	8.74E+32	&	1.48E+34	&	1.48	&	0.0872	&	\citet{Folsom2016}	\\
HIP 12545	&	73.9	&	40.8	&	32.7	&	8.07E-14	&	2.53E-11	&	6.49E-14	&	7.89E-10	&	1.96E+32	&	4.41E+33	&	1.56	&	0.0693	&	\citet{Folsom2016}	\\
HIP 76768	&	68.2	&	18.6	&	20.4	&	5.01E-14	&	2.34E-11	&	6.01E-14	&	7.21E-10	&	9.24E+31	&	2.59E+33	&	3.17	&	0.113	&	\citet{Folsom2016}	\\
Lo Peg	&	81.3	&	43.1	&	37.3	&	4.43E-14	&	1.63E-11	&	4.19E-14	&	3.34E-10	&	4.07E+32	&	1.00E+34	&	5.6	&	0.228	&	\citet{Folsom2016}	\\
PELS 031	&	13.4	&	11	&	11.1	&	4.99E-13	&	1.67E-10	&	1.16E-12	&	1.60E-11	&	2.01E+32	&	8.03E+33	&	2.94	&	0.0734	&	\citet{Folsom2016}	\\
PW And	&	102	&	44.5	&	27	&	1.79E-13	&	1.30E-11	&	3.33E-14	&	8.97E-10	&	4.21E+32	&	4.29E+33	&	1.63	&	0.16	&	\citet{Folsom2016}	\\
TYC 0486-4943-1	&	14.9	&	12.4	&	10.5	&	4.15E-14	&	9.83E-11	&	7.33E-13	&	7.55E-12	&	1.08E+31	&	1.29E+33	&	23.9	&	0.198	&	\citet{Folsom2016}	\\
TYC 5164-567-1	&	51.6	&	20.4	&	18.1	&	3.15E-13	&	4.39E-11	&	1.13E-13	&	3.36E-10	&	1.72E+32	&	2.50E+33	&	1.66	&	0.114	&	\citet{Folsom2016}	\\
TYC 6349-0200-1	&	48.6	&	17	&	13.8	&	4.64E-14	&	4.47E-11	&	1.15E-13	&	4.06E-10	&	1.02E+32	&	4.22E+33	&	3.49	&	0.0843	&	\citet{Folsom2016}	\\
TYC 6878-0195-1	&	43.5	&	16.7	&	12	&	3.68E-13	&	8.03E-11	&	2.06E-13	&	6.32E-10	&	4.81E+32	&	8.92E+33	&	0.766	&	0.0414	&	\citet{Folsom2016}	\\
HD 6569	&	16.7	&	5.12	&	3.43	&	1.12E-13	&	5.23E-11	&	1.65E-13	&	2.93E-11	&	1.41E+31	&	4.42E+32	&	12	&	0.385	&	\citet{Folsom2018}	\\
HIP 10272	&	11	&	3.94	&	1.78	&	2.16E-13	&	7.69E-11	&	4.31E-13	&	1.26E-11	&	1.86E+31	&	6.86E+32	&	11.7	&	0.319	&	\citet{Folsom2018}	\\
BD-072388	&	84.5	&	41.2	&	38.4	&	2.48E-13	&	1.78E-11	&	4.57E-14	&	5.58E-10	&	2.28E+33	&	2.32E+34	&	1.63	&	0.16	&	\citet{Folsom2018}	\\
HD 141943 (2007)	&	29.4	&	23.8	&	29.3	&	2.54E-11	&	1.77E-10	&	4.55E-13	&	1.89E-10	&	1.39E+34	&	3.97E+34	&	8.66E-02	&	0.0302	&	\citet{Marsden2011}	\\
HD 141943 (2009)	&	19.5	&	16.4	&	9.5	&	2.54E-11	&	2.40E-10	&	9.09E-13	&	7.96E-11	&	9.52E+33	&	3.97E+34	&	0.126	&	0.0302	&	\citet{Marsden2011}	\\
HD 141943 (2010)	&	14.9	&	17.4	&	28.1	&	2.54E-11	&	2.35E-10	&	1.43E-12	&	2.91E-11	&	7.43E+33	&	3.97E+34	&	0.161	&	0.0302	&	\citet{Marsden2011}	\\
HD 35296 (2007)	&	6.69	&	3.66	&	4.78	&	3.56E-12	&	2.70E-11	&	1.31E-13	&	6.61E-12	&	2.49E+32	&	1.05E+33	&	2.05	&	0.484	&	\citet{Waite2015}	\\
HD 35296 (2008)	&	3	&	4.76	&	4.6	&	3.56E-12	&	2.94E-11	&	4.48E-13	&	3.10E-13	&	1.90E+32	&	1.05E+33	&	2.69	&	0.484	&	\citet{Waite2015}	\\
HD 29615	&	60.6	&	26.1	&	28.6	&	3.77E-12	&	3.56E-11	&	9.13E-14	&	5.63E-10	&	2.19E+33	&	7.38E+33	&	0.288	&	0.0854	&	\citet{Waite2015}	\\
EK Dra (2006)	&	25.9	&	18.6	&	16.6	&	1.40E-12	&	1.14E-10	&	3.85E-13	&	5.26E-11	&	4.09E+32	&	5.15E+33	&	1.3	&	0.104	&	\citet{Waite2017}	\\
EK Dra (Jan 2007)	&	33.8	&	19.8	&	13.5	&	1.40E-12	&	9.59E-11	&	2.46E-13	&	1.12E-10	&	5.21E+32	&	5.15E+33	&	1.02	&	0.104	&	\citet{Waite2017}	\\
EK Dra (Feb 2007)	&	14.9	&	12	&	13.6	&	1.40E-12	&	1.45E-10	&	9.88E-13	&	1.49E-11	&	2.45E+32	&	5.15E+33	&	2.17	&	0.104	&	\citet{Waite2017}	\\
EK Dra (2008)	&	21.5	&	10.7	&	12.7	&	1.40E-12	&	1.39E-10	&	5.28E-13	&	5.36E-11	&	3.44E+32	&	5.15E+33	&	1.55	&	0.104	&	\citet{Waite2017}	\\
EK Dra (2012)	&	13.3	&	30.7	&	23.4	&	1.40E-12	&	1.15E-10	&	1.19E-12	&	2.19E-12	&	2.22E+32	&	5.15E+33	&	2.4	&	0.104	&	\citet{Waite2017}	\\
																									
\textbf{Hot Jupiter Hosts} \\																									
$\tau$ Boo (Jan 2008)	&	0.889	&	0.794	&	0.873	&	7.66E-12	&	5.71E-11	&	1.14E-12	&	1.19E-13	&	4.54E+32	&	2.39E+33	&	2.17	&	0.413	&	\citet{Fares2009}	\\
$\tau$ Boo (Jun 08)	&	0.868	&	0.778	&	0.772	&	7.66E-12	&	5.84E-11	&	1.17E-12	&	1.13E-13	&	4.45E+32	&	2.39E+33	&	2.21	&	0.413	&	\citet{Fares2009}	\\
$\tau$ Boo (Jul 2008)	&	0.725	&	0.784	&	0.696	&	7.66E-12	&	6.07E-11	&	1.21E-12	&	6.06E-14	&	4.31E+32	&	2.39E+33	&	2.28	&	0.413	&	\citet{Fares2009}	\\
$\tau$ Boo (2009)	&	1.23	&	1	&	0.831	&	7.66E-12	&	5.29E-11	&	1.03E-12	&	2.58E-13	&	4.84E+32	&	2.39E+33	&	2.03	&	0.413	&	\citet{Fares2013}	\\
$\tau$ Boo (2010)	&	1.35	&	1.19	&	1.04	&	7.66E-12	&	4.95E-11	&	9.36E-13	&	2.80E-13	&	5.11E+32	&	2.39E+33	&	1.93	&	0.413	&	\citet{Fares2013}	\\
$\tau$ Boo (Jan 2011)	&	1.63	&	0.745	&	0.899	&	7.66E-12	&	5.14E-11	&	9.86E-13	&	8.37E-13	&	4.95E+32	&	2.39E+33	&	1.99	&	0.413	&	\citet{Fares2013}	\\
$\tau$ Boo (May 2011)	&	0.742	&	1.23	&	1.25	&	7.66E-12	&	5.18E-11	&	1.03E-12	&	3.18E-14	&	4.92E+32	&	2.39E+33	&	2	&	0.413	&	\citet{Mengel2016}	\\
\hline
\end{tabular}
\end{sidewaystable}

\begin{sidewaystable}
\contcaption{continued}
\center
\begin{tabular}{lcccccccccccc}
\hline\hline
Star ID & $\langle B_{\rm d} \rangle$ & $\langle B_{\rm q} \rangle$ & $\langle B_{\rm o} \rangle$ & $\dot{M}$ (CS11) & $\dot{M}$ (M15) & $\dot{M}$ (mod M15) & $\dot{M}_{\rm crit}$ & $T$ (CS11) & $T$ (M15) & $\tau$ (CS11) & $\tau$ (M15) & Reference\\
 & (G) & (G) & (G) & ($M_{\odot}\ {\rm yr}^{-1}$) & ($M_{\odot}\ {\rm yr}^{-1}$) & ($M_{\odot}\ {\rm yr}^{-1}$) & ($M_{\odot}\ {\rm yr}^{-1}$) & (erg) & (erg) & (Gyr) & (Gyr) & \\
\hline
$\tau$ Boo (May 2013)	&	1.32	&	1.26	&	1.08	&	7.66E-12	&	4.91E-11	&	9.28E-13	&	2.41E-13	&	5.14E+32	&	2.39E+33	&	1.91	&	0.413	&	\citet{Mengel2016}	\\
$\tau$ Boo (Dec 2013)	&	1.92	&	1.74	&	0.925	&	7.66E-12	&	4.46E-11	&	7.18E-13	&	5.49E-13	&	5.56E+32	&	2.39E+33	&	1.77	&	0.413	&	\citet{Mengel2016}	\\
$\tau$ Boo (2014)	&	1.05	&	0.893	&	0.965	&	7.66E-12	&	5.41E-11	&	1.08E-12	&	1.75E-13	&	4.75E+32	&	2.39E+33	&	2.07	&	0.413	&	\citet{Mengel2016}	\\
$\tau$ Boo (Jan 2015)	&	1.14	&	1.02	&	1.06	&	7.66E-12	&	5.18E-11	&	1.03E-12	&	1.96E-13	&	4.92E+32	&	2.39E+33	&	2	&	0.413	&	\citet{Mengel2016}	\\
$\tau$ Boo (2 Apr 2015)	&	1.07	&	0.435	&	0.309	&	7.66E-12	&	6.60E-11	&	1.32E-12	&	3.95E-13	&	4.03E+32	&	2.39E+33	&	2.44	&	0.413	&	\citet{Mengel2016}	\\
$\tau$ Boo (13 Apr 2015)	&	0.826	&	0.378	&	0.229	&	7.66E-12	&	7.28E-11	&	1.45E-12	&	2.15E-13	&	3.71E+32	&	2.39E+33	&	2.65	&	0.413	&	\citet{Mengel2016}	\\
$\tau$ Boo (20 Apr 2015)	&	1	&	0.426	&	0.321	&	7.66E-12	&	6.70E-11	&	1.34E-12	&	3.37E-13	&	3.98E+32	&	2.39E+33	&	2.47	&	0.413	&	\citet{Mengel2016}	\\
$\tau$ Boo (May 2015)	&	1.34	&	0.811	&	0.745	&	7.66E-12	&	5.41E-11	&	1.06E-12	&	4.35E-13	&	4.75E+32	&	2.39E+33	&	2.07	&	0.413	&	\citet{Mengel2016}	\\
HD 73256	&	3.54	&	3.07	&	3.44	&	9.98E-14	&	7.89E-12	&	5.97E-14	&	6.41E-13	&	2.56E+30	&	4.91E+31	&	48.8	&	2.54	&	\citet{Fares2013}	\\
HD 102195	&	6.62	&	3.37	&	2.45	&	7.87E-14	&	1.59E-11	&	7.30E-14	&	3.67E-12	&	3.67E+30	&	8.94E+31	&	28	&	1.15	&	\citet{Fares2013}	\\
HD 130322	&	1.82	&	1.08	&	0.451	&	1.72E-14	&	4.01E-12	&	3.27E-14	&	2.60E-13	&	2.47E+29	&	8.90E+30	&	164	&	4.56	&	\citet{Fares2013}	\\
HD 179949 (2007)	&	1.17	&	1.16	&	1.07	&	5.10E-13	&	9.09E-12	&	1.53E-13	&	1.12E-13	&	1.33E+31	&	1.40E+32	&	22.1	&	2.1	&	\citet{Fares2012}	\\
HD 179949 (2009)	&	2.04	&	1.94	&	1.29	&	5.10E-13	&	7.55E-12	&	6.60E-14	&	3.67E-13	&	1.78E+31	&	1.40E+32	&	16.6	&	2.1	&	\citet{Fares2012}	\\
HD 189733 (2007)	&	3.88	&	7.53	&	7	&	3.94E-14	&	1.71E-11	&	2.44E-13	&	1.64E-13	&	1.22E+30	&	8.94E+31	&	74.7	&	1.02	&	\citet{Fares2010}	\\
HD 189733 (2008)	&	9.17	&	8.26	&	5.71	&	3.94E-14	&	1.48E-11	&	6.57E-14	&	3.13E-12	&	2.67E+30	&	8.94E+31	&	34	&	1.02	&	\citet{Fares2010}	\\
																									
\textbf{M dwarf Stars} \\																									
CE Boo	&	98.7	&	25	&	17.9	&	9.46E-17	&	1.61E-12	&	4.14E-15	&	3.69E-10	&	1.44E+29	&	2.82E+31	&	180	&	0.916	&	\citet{Donati2008}	\\
DS Leo (2007)	&	33.1	&	27.5	&	12.2	&	2.72E-16	&	1.35E-11	&	3.46E-14	&	2.08E-11	&	1.71E+29	&	5.99E+31	&	244	&	0.696	&	\citet{Donati2008}	\\
DS Leo (2008)	&	32.6	&	24.3	&	11.6	&	2.72E-16	&	1.38E-11	&	3.55E-14	&	2.31E-11	&	1.68E+29	&	5.99E+31	&	247	&	0.696	&	\citet{Donati2008}	\\
GJ 182	&	72.6	&	44.3	&	40.1	&	3.36E-15	&	1.94E-11	&	4.98E-14	&	3.96E-10	&	1.74E+31	&	1.91E+33	&	12.7	&	0.116	&	\citet{Donati2008}	\\
GJ 49	&	15.7	&	8.71	&	5.32	&	1.82E-16	&	1.25E-11	&	3.91E-14	&	7.11E-12	&	4.93E+28	&	2.29E+31	&	612	&	1.32	&	\citet{Donati2008}	\\
AD Leo (2007)	&	162	&	74.6	&	48.6	&	6.12E-17	&	2.44E-12	&	6.26E-15	&	5.09E-10	&	8.20E+29	&	2.55E+32	&	152	&	0.489	&	\citet{Morin2008}	\\
AD Leo (2008)	&	170	&	64.5	&	39.3	&	6.12E-17	&	2.25E-12	&	5.78E-15	&	6.57E-10	&	8.56E+29	&	2.55E+32	&	146	&	0.489	&	\citet{Morin2008}	\\
DT Vir (2007)	&	79.3	&	40.6	&	26.5	&	5.98E-16	&	1.25E-11	&	3.22E-14	&	2.14E-10	&	3.03E+30	&	6.67E+32	&	70.1	&	0.319	&	\citet{Donati2008}	\\
DT Vir (2008)	&	39	&	48.5	&	47.9	&	5.98E-16	&	3.30E-11	&	1.07E-13	&	1.69E-11	&	1.58E+30	&	6.67E+32	&	134	&	0.319	&	\citet{Donati2008}	\\
EQ Peg A	&	366	&	107	&	114	&	4.78E-17	&	5.63E-13	&	1.44E-15	&	3.09E-09	&	2.50E+30	&	4.03E+32	&	87.2	&	0.542	&	\citet{Morin2008}	\\
EQ Peg B	&	379	&	115	&	96	&	1.73E-18	&	2.99E-13	&	7.68E-16	&	1.74E-09	&	4.37E+29	&	3.01E+32	&	435	&	0.631	&	\citet{Morin2008}	\\
EV Lac (2006)	&	447	&	213	&	102	&	1.80E-17	&	3.11E-13	&	7.99E-16	&	2.37E-09	&	2.77E+29	&	5.49E+31	&	113	&	0.57	&	\citet{Morin2008}	\\
EV Lac (2007)	&	420	&	164	&	91.5	&	1.80E-17	&	3.45E-13	&	8.86E-16	&	2.48E-09	&	2.62E+29	&	5.49E+31	&	119	&	0.57	&	\citet{Morin2008}	\\
DX Cnc (2007)	&	92.3	&	37.2	&	33.8	&	4.54E-21	&	1.02E-12	&	2.61E-15	&	1.70E-11	&	3.87E+26	&	1.30E+31	&	4.13E+04	&	1.23	&	\citet{Morin2010}	\\
DX Cnc (2008)	&	45.3	&	42.2	&	27.1	&	4.54E-21	&	2.82E-12	&	8.68E-15	&	1.67E-12	&	2.02E+26	&	1.30E+31	&	7.93E+04	&	1.23	&	\citet{Morin2010}	\\
DX Cnc (2009)	&	53.5	&	26.4	&	17	&	4.54E-21	&	2.55E-12	&	6.55E-15	&	4.81E-12	&	2.35E+26	&	1.30E+31	&	6.80E+04	&	1.23	&	\citet{Morin2010}	\\
GJ 1156 (2007)	&	35.3	&	31.9	&	26.5	&	2.92E-19	&	5.70E-12	&	2.02E-14	&	2.27E-12	&	4.33E+27	&	4.61E+31	&	8150	&	0.765	&	\citet{Morin2010}	\\
GJ 1156 (2008)	&	74.9	&	62	&	40.1	&	2.92E-19	&	2.21E-12	&	5.66E-15	&	1.15E-11	&	8.63E+27	&	4.61E+31	&	4090	&	0.765	&	\citet{Morin2010}	\\
GJ 1156 (2009)	&	71.4	&	47.7	&	30.7	&	2.92E-19	&	2.39E-12	&	6.14E-15	&	1.35E-11	&	8.26E+27	&	4.61E+31	&	4270	&	0.765	&	\citet{Morin2010}	\\
GJ 1245 B (2006)	&	122	&	69.4	&	63.1	&	1.09E-19	&	7.98E-13	&	2.05E-15	&	3.60E-11	&	3.71E+27	&	1.95E+31	&	4410	&	0.841	&	\citet{Morin2010}	\\
GJ 1245 B (2008)	&	38.2	&	33.6	&	28.5	&	1.09E-19	&	4.39E-12	&	1.45E-14	&	2.14E-12	&	1.28E+27	&	1.95E+31	&	1.28E+04	&	0.841	&	\citet{Morin2010}	\\
OT Ser	&	66.3	&	47.6	&	53.4	&	2.82E-16	&	1.56E-11	&	4.00E-14	&	8.85E-11	&	1.14E+30	&	4.23E+32	&	134	&	0.361	&	\citet{Donati2008}	\\
V374 Peg (2005)	&	588	&	235	&	183	&	4.92E-18	&	1.64E-13	&	4.22E-16	&	4.30E-09	&	1.42E+30	&	4.03E+32	&	157	&	0.556	&	\citet{Morin2008V374Peg}	\\
V374 Peg (2006)	&	491	&	243	&	124	&	4.92E-18	&	2.23E-13	&	5.71E-16	&	2.50E-09	&	1.21E+30	&	4.03E+32	&	186	&	0.556	&	\citet{Morin2008V374Peg}	\\
WX UMa (2006)	&	904	&	511	&	275	&	1.28E-20	&	2.14E-14	&	5.48E-17	&	1.49E-09	&	4.25E+27	&	1.00E+31	&	2.22E+03	&	0.94	&	\citet{Morin2010}	\\
WX UMa (2007)	&	1160	&	485	&	219	&	1.28E-20	&	1.41E-14	&	3.61E-17	&	3.22E-09	&	5.33E+27	&	1.00E+31	&	1.77E+03	&	0.94	&	\citet{Morin2010}	\\
WX UMa (2008)	&	1110	&	593	&	326	&	1.28E-20	&	1.50E-14	&	3.85E-17	&	2.41E-09	&	5.15E+27	&	1.00E+31	&	1.83E+03	&	0.94	&	\citet{Morin2010}	\\
WX UMa (2009)	&	1590	&	212	&	229	&	1.28E-20	&	8.23E-15	&	2.11E-17	&	1.15E-08	&	7.13E+27	&	1.00E+31	&	1.32E+03	&	0.94	&	\citet{Morin2010}	\\
YZ CMi (2007)	&	540	&	205	&	146	&	1.47E-17	&	2.27E-13	&	5.81E-16	&	3.86E-09	&	4.18E+29	&	7.80E+31	&	118	&	0.633	&	\citet{Morin2008}	\\
YZ CMi (2008)	&	514	&	185	&	166	&	1.47E-17	&	2.47E-13	&	6.33E-16	&	3.64E-09	&	4.00E+29	&	7.80E+31	&	123	&	0.633	&	\citet{Morin2008}	\\
GJ 176	&	6.6	&	15.4	&	14.2	&	9.04E-18	&	2.13E-12	&	1.46E-14	&	1.30E-13	&	1.66E+27	&	2.31E+30	&	6.12E+03	&	4.4	&	H\'{e}brard et al. (in prep)	\\
GJ 205	&	18.8	&	5.22	&	2.7	&	6.44E-17	&	1.41E-12	&	3.61E-15	&	2.06E-11	&	2.28E+28	&	5.12E+30	&	903	&	4.01	&	\citet{Hebrard2016}	\\
GJ 358	&	123	&	15.9	&	8.15	&	1.22E-17	&	1.96E-13	&	5.03E-16	&	7.25E-10	&	2.96E+28	&	5.65E+30	&	372	&	1.95	&	\citet{Hebrard2016}	\\
GJ 479	&	31.4	&	13.7	&	11	&	1.81E-17	&	2.32E-12	&	5.95E-15	&	2.54E-11	&	1.20E+28	&	7.02E+30	&	1.02E+03	&	1.75	&	\citet{Hebrard2016}	\\
GJ 674	&	119	&	11.9	&	5.21	&	1.16E-18	&	8.00E-14	&	2.05E-16	&	7.55E-10	&	5.60E+27	&	2.35E+30	&	856	&	2.04	&	H\'{e}brard et al. (in prep)	\\
GJ 846 (2013)	&	9.09	&	2.1	&	1.08	&	4.96E-16	&	8.01E-11	&	7.81E-13	&	5.21E-12	&	1.06E+29	&	1.45E+32	&	554	&	0.404	&	\citet{Hebrard2016}	\\
GJ 846 (2014)	&	18.2	&	7.29	&	2.42	&	4.96E-16	&	5.67E-11	&	2.43E-13	&	1.44E-11	&	1.99E+29	&	1.45E+32	&	294	&	0.404	&	\citet{Hebrard2016}	\\
\hline
\end{tabular}
\end{sidewaystable}

\end{document}